\documentclass[a4paper,11pt]{article}
\usepackage[pdftex]{graphicx}
\usepackage[T1]{fontenc}
\usepackage{lmodern}
\usepackage{slashed}
\usepackage[utf8]{inputenc}
\usepackage[english]{babel}
\usepackage{microtype}
\usepackage{cite}
\usepackage{amsmath}
\usepackage{amssymb}
\usepackage{amsfonts}
\usepackage[nottoc,notlot,notlof]{tocbibind}
\usepackage{upgreek}
\usepackage{mathtools}
\numberwithin{equation}{section}
\allowdisplaybreaks
\usepackage[all]{xy}
\usepackage{color}
\usepackage{graphicx}
\graphicspath{{images/}}
\usepackage{geometry}
\usepackage[toc,page]{appendix}
\usepackage{hyperref}
\usepackage[normalem]{ulem}

\newcommand{\diff}{\mathrm{d}}

\usepackage{bm}
\usepackage{ragged2e}
\usepackage{appendix}
\usepackage{slashed}
\usepackage{bbold}
\usepackage{cancel}
\usepackage{multirow}
\usepackage{array,multirow}
\usepackage{multicol}
\usepackage{multirow}
\usepackage{array}
    \newcolumntype{P}[1]{>{\centering\arraybackslash}p{#1}}
    \newcolumntype{M}[1]{>{\centering\arraybackslash}m{#1}}

\usepackage{amsthm}

\newtheorem*{theorem*}{Theorem}

\DeclareMathAlphabet{\mathpzc}{OT1}{pzc}{m}{it}

\definecolor{blue-violet}{rgb}{0.54, 0.17, 0.89}
\definecolor{PineGreen}{cmyk}{0.92, 0, 0.59, 0.25}
\definecolor{YellowOrange}{cmyk}{0, 0.42, 1, 0}
\definecolor{orange}{rgb}{0.95, 0.5, 0.1}

\usepackage[dvipsnames]{xcolor}

\usepackage{caption} 
\usepackage{float}

\begin{document}

\begin{titlepage}
\begin{flushright}
\par\end{flushright}
\vskip 0.5cm
\begin{center}
\textbf{\Huge \bf Supergravities and Branes}\\
\vskip .2cm
\textbf{\Huge \bf from Hilbert-Poincaré Series}

\vskip 1cm 

\large {\bf C.~A.~Cremonini}$^{~a,}$\footnote{carlo.alberto.cremonini@gmail.com} 
\large {\bf P.~A.~Grassi}$^{~b, c, }$\footnote{pietro.grassi@uniupo.it},
\large {\bf R.~Noris}$^{~d,}$\footnote{noris@fzu.cz},
\large {\bf L.~Ravera}$^{~e, c,}$\footnote{lucrezia.ravera@polito.it},

\vskip .5cm {
\small
{$^{(a)}$ \it Faculty of Mathematics and Physics, Mathematical Institute, Charles University Prague, Sokolovska 49/83, 186 75 Prague}\\
{$^{(b)}$ \it DiSIT, Universit\`a del Piemonte Orientale, Viale T. Michel 11, 15121 Alessandria, Italy}\\
{$^{(c)}$ \it INFN, Sezione di Torino, Via P. Giuria 1, 10125 Torino, Italy} \\
{$^{(d)}$ \it CEICO, Institute of Physics of the Czech Academy of Sciences, \\
Na Slovance 2, 182 21 Prague 8, Czech Republic}\\
{$^{(e)}$ \it Politecnico di Torino, Corso Duca degli Abruzzi, 24, 10129 Torino, Italy}
}
\end{center}

\begin{abstract} 
The Molien-Weyl integral formula and the Hilbert-Poincaré series have proven to be powerful mathematical tools in relation to gauge theories, allowing to count the number of gauge invariant operators. In this paper we show that these methods can also be employed to construct Free Differential Algebras and, therefore, reproduce the associated pure supergravity spectrum and nonperturbative objects.
Indeed, given a set of fields, the Hilbert-Poincaré series allows to compute all possible invariants and consequently derive the cohomology structure.
\end{abstract}

\vfill{}
\vspace{1.5cm}
\end{titlepage}

\setcounter{footnote}{0}
\tableofcontents

\section{Introduction}

Supergravity, being the supersymmetric extension of Einstein's general relativity, is deeply connected to geometry and their relation has been explored in several directions. A particularly important role is played by superspaces and supermanifolds, whose properties can be described in terms of the usual vielbein $V^a$ and its supersymmetric partner $\psi^\alpha$, which can be merged in a single object, the supervielbein $\mathcal{V}^A = (V^a, \psi^\alpha)$. The basic geometric structure can be completed by including the spin connection $\omega^{[ab]}$, the gauge field associated with the local Lorentz group, the latter playing an important role in the forthcoming sections.\\

To define meaningful supergravity models with a suitable amount of supersymmetry, one has to specify the spectrum of necessary degrees of freedom (d.o.f.) for the quantum field theory. To this end, a very constructive technique is that of \emph{Free Differential Algebras} (FDAs), discussed in several books and papers \cite{DFd11,gm13,gm3,CDF} and based on the mathematical construction by D. Sullivan \cite{sullivan}. \\
As we will review in more detail in the main text, the construction of a FDA {starts} by considering a Lie superalgebra $\mathfrak{g}$, associated with a Lie supergroup $G$ and described in terms of its structure constants $f^{A}_{\phantom{A} BC}$. The algebraic structure can be equivalently expressed in  terms of its dual Maurer-Cartan (MC) equations.\\
In particular, for supergravity theories, one typically considers a subalgebra $\mathfrak{h}$ reductive in the ambient Lie algebra $\mathfrak{g}$, in such a way that the one forms dual to the generators of translations and supersymmetries can be identified with the supervielbein of the coset $G/H$ and satisfy 
\begin{eqnarray}
\label{introA}
\diff \mathcal{V}^A = {-\frac 12} {f^{A}}_{BC} \mathcal{V}^B \wedge \mathcal{V}^C\,, ~~~~~ \diff^2 =0 \,.
\end{eqnarray}
The latter describe the vacuum structure of the supergravity theory both for flat superspaces and for curved rigid supermanifolds.
Out of the vacuum these provide the building blocks (i.e., the curvatures) for the study of dynamics.\\
In this framework, $\diff$ is a nilpotent operator and one can study its cohomology {$H^\bullet \left( \mathfrak{g} , \mathfrak{h} ; \mathbb C \right)$}, {also known as} the Chevalley-Eilenberg (CE) cohomology \cite{CE,Fuks}, on the space of super differential forms (hereafter named {\it superforms}) $\Omega^{\bullet}( \mathfrak{g} , \mathfrak{h} ; \mathbb C )$ with constant coefficients. When $H^\bullet \left( \mathfrak{g} , \mathfrak{h} ; \mathbb C \right)$ is not empty, one can introduce new superforms $A^{(p)}$ whose differentials are exactly the cocycles in $H^\bullet \left( \mathfrak{g} , \mathfrak{h} ; \mathbb C \right)$.
One can repeat this procedure on complex valued polynomials on $\mathcal{V}^A$ and $A^{(p)}$ and, by iterating the construction, one has two possible outcomes: either the algorithm ends in a finite number of steps or one has to introduce an infinite number of $p$-forms. \\
The obtained forms $A^{(p)}$ can be interpreted as new potentials to be added to the theory in order to complete the spectrum. 
In diverse contexts, the full set of superforms, together with their exterior derivatives, has been shown to describe the vacuum configuration of the given supergravity theory.
In such cases, their dynamics can be derived from a suitably constructed Lagrangian \cite{CDF}. \\

The study of FDA cohomologies may involve the presence of several forms in different representations and can lead to unhandleable problems by brute force computations. In the present work, we make use of a technique, based on the \emph{Molien-Weyl formula} \cite{DerksenKemper,Hanany:2008sb,procesi}, to compute invariant polynomials and cohomology groups in various cases, associated with the local description of both flat and curved rigid supermanifolds. \\
The Molien-Weyl formula is a powerful tool which allows {to compute the number of} invariant form polynomials in terms of the Hilbert-Poincar\'e series $P(t) = \sum_{n} b_n t^n$. As we will discuss in the main text, the parameter $t$ is associated with a chosen scaling of the fields (form degree, weight under scaling, $R$-symmetry weight, etc.), whereas the coefficients $b_n$ count the number of invariants at fixed $t$, that is, $b_n$ will be then related to the dimension of cohomology groups, analogously to Betti numbers in the de Rham case.
\\

The invariant polynomials computed in this way are written in terms of the supervielbein $\mathcal{V}^A = (V^a, \psi^\alpha)$, whose components respectively transform under the vector and spinor representations of the Lorentz group, and possibly of the added superforms $A^{(p)}$. Moreover, in the case of extended supergravities, one should add more labels to the spinor 1-forms by considering representations of the $R$-symmetry group. 
At last, the Molien-Weyl formula also takes into account the commuting/anticommuting nature of the d.o.f.: as an example, the vielbeins are anticommuting one forms (with respect to the wedge product), while the gravitini $\psi^\alpha$ are commuting ones. \\

The paper is organized as follows: In Sec. \ref{sec2} we introduce the necessary mathematical tools and discuss their use in relation to various vacuum supergravity theories. In particular, we consider $D=4$ (Sec. \ref{sec3}), $D=6$ (Sec. \ref{sec4}), $D=10$ (Sec. \ref{sec5}), $D=11$ (Sec. \ref{sec6}) and $D=12$ (Sec. \ref{sec7}) flat and curved spacetimes, with different amounts of supersymmetry (labeled by $N$).\\
In Sec. \ref{sec8}, we instead discuss the application of the aforementioned techniques to the study of filtered deformations, which can be put in relations with the Killing superalgebra of supergravity backgrounds. At last, Sec. \ref{sec9} is devoted to a final discussion and future developments. 

\section{Theoretical framework}\label{sec2}

\subsection{Molien-Weyl formula}

Let $G$ be a finite group and $\rho_R$ a representation on a finite dimensional vector space $V$ (over $\mathbb{R}$ or $\mathbb{C}$), namely for each $g\in G$, $\rho_R(g)$ acts on $V$ as a matrix in the representation $R$. \\
Through the use of the Hilbert-Poincar\'e series, the $G$-invariant polynomials are computed by means of the Molien formula (see, e.g., \cite{DerksenKemper,procesi,Pouliot:1998yv,Weyl}) 
\begin{align}\label{MW_A}
P( V^G, t) = \frac{1}{|G|} \sum_{g \in G} \frac{1}{\det(1 - t \rho_R(g))} = \sum_{n\geq 0} b_n t^n \,,
\end{align}
where $V^G$ {is the vector space of} the $G$-invariant polynomials and $b_n$ {the dimension of the subspaces of invariant polynomials at order $n$.} 
Let us now consider a graded 
vector space $V = V_0 \oplus V_1$, where $V_0$ is the bosonic subspace and $V_1$ is the fermionic one, with associated representations $\rho_0$ and 
$\rho_1$. The Molien formula can then be rewritten as follows: 
\begin{align}\label{MW_B}
P( V^G, u, t) = \frac{1}{|G|} \sum_{g \in G} \frac{\det (1 - u \rho_1(g))}{\det (1 - t \rho_0(g))} \,,
\end{align}
where we introduced two different parameters $u$ and $t$ to parametrize the representantions $\rho_0$ and $\rho_1$ {in the $V_0$ and $V_1$ subspace, respectively}.  Interestingly, the sum in \eqref{MW_B} can actually be restricted to the conjugacy classes of the finite group $G$.\\

If we consider a continuous Lie group, instead of a  finite group, the sum is replaced by an integral as follows:
\begin{align}\label{MW_C}
P( V^G, u, t) = \int_{{G}_c} \frac{\det (1 - u \rho_1(g))}{\det (1 - t \rho_0(g))} \diff \mu_G \,,
 \end{align}
where ${G}_c$ is the maximal compact subgroup of $G$ and $\diff \mu_G$ is the corresponding Haar measure normalized such that $\int _{{G}_c} \diff \mu_G=1$. Since the integrand is invariant under conjugation, one should only integrate on the abelian Cartan subgroup: this is the content of the Weyl integration formula \cite{procesi,Weyl}, which allows to rewrite the above formula as
\begin{align}\label{MW_D0}
P( V^G, u, t) = \int_{T_c} \frac{\det (1 - u \rho_1(g_c))}{\det (1 - t \rho_0(g_c))} \phi(g_c)  \diff \nu_c \,,
\end{align}
where the integration is restricted to $g_c$, namely to the elements of the abelian Cartan subgroup $T_c$. {Notice that this step requires the restriction to the component connected to the identity of the chosen Lie group $G$.} The measure $\diff \mu_G  = \phi(g_c)  \diff \nu_c$ is split into $\phi(g_c)$, which is written in terms of the positive roots of $G$, and $\diff \nu_c$, the integration measure over the Cartan subgroup. The integration over the Cartan subgroup can be performed by introducing a complex coordinate $z_i$ ($i=1, \ldots, {\rm rk}\, G$), defined on the unit circle $|z_i|=1$, for each Cartan generator. Consequently, by using the residue theorem one obtains
\begin{align}\label{MW_D}
P( V^G, u, t) = \oint_{|z_i|=1} \frac{\det (1 - u \rho_1(g_c(z_i)))}{\det (1 - t \rho_0(g_c(z_i)))} \phi(z_i) \prod_{i=1}^{{\rm rk}\, G} 
\frac{\diff z_i}{2 \pi i z_i} \,,
\end{align}
with
\begin{align}
\label{MW_E}
\phi(z_i)  = \prod_{\vec{\alpha} \in \Delta^+} \left( 1 - \prod_{i=1}^{{\rm rk}\, G} z_i^{\alpha_i} \right) \,, 
\end{align}
{where $\vec{\alpha}$ represents an element of the finite set of positive roots $\Delta^+$ and $\alpha_i$ are the components of $\vec{\alpha}$ in terms of the Cartan generators in the Chevalley basis.} 
Concerning the integrand, we use the following notion: given a certain representation $\rho_{R}$ of the group, we can express the character of the representation as
\begin{align}\label{MW_F} 
{\chi_R(z_i) = Tr\left[ \rho_R(g_c(z_i))\right]} \,,
\end{align}
where $g_c(z_i)$ is an element of the Cartan subgroup $G_c$. For finite dimensional representations, $\chi_R(z_i)$ can be expressed as the sum of monomials $m(\vec{\lambda}; z_i) = \prod_{i=1}^{{\rm rk}\, G} z_i^{\lambda_i}$, where $\lambda_i\in\Delta_R$ are the weights of the representation $R$ (we take into account both vanishing and nonvanishing ones), in the following way:
\begin{align}\label{MW_FA}
\chi_R(z_i) = \sum_{\vec{\lambda} \in \Delta_R} \prod_{i=1}^{{\rm rk}\, G} z_i^{\lambda_i} = 
 \sum_{\vec{\lambda} \in \Delta_R} m(\vec{\lambda}; z_i) \,. 
\end{align}
One can then construct the \emph{plethystic exponential} as follows (the subscript ``$B$'' indicates that we are dealing with bosonic variables): 
\begin{align}\label{MW_G}
PE_B[\chi_R(z_i) t] = Exp \left[ - \sum_{\vec{\lambda} \in \Delta_R}  
\sum_{n=0}^\infty \frac 1n t^n \prod_{i=1}^{{\rm rk}\, G} z_i^{ n \lambda_i} \right] = 
\frac{1}{\prod_{\vec{\lambda} \in \Delta_R} (1 - t \, m(\vec{\lambda}; z_i))} \,,
\end{align}
corresponding to the denominator in \eqref{MW_D},
\begin{align}\label{MW_H}
PE_B[\chi_R(z_i) t] =  \frac{1}{\det (1 - t \rho_R(g_c(z_i)))} \,.
\end{align}
This formula is suitable for a conventional vector space and its representations. However, 
for fermionic(anticommuting) d.o.f. it has to be modified as (see \cite{Feng:2007ur})
\begin{align}\label{MW_J}
{PE}_F[\chi_R(z_i) t] = Exp \left[ \sum_{\vec{\lambda} \in \Delta_R}  
\sum_{n=0}^\infty \frac 1n t^n \prod_{i=1}^{{\rm rk}\, G} z_i^{ n \lambda_i} \right] = 
\prod_{\vec{\lambda} \in \Delta_R} (1 - t \, m(\vec{\lambda}; z_i))= \det (1 - t\rho_R(g_c(z_i))) \,,
\end{align}
where the minus sign in the definition of the plethystic exponential \eqref{MW_H} has been removed. \\
In general, one can add different representations $R = \oplus_{I=1}^N R_I$ and, by distinguishing the statistic nature of the vector space, {the integrand of \eqref{MW_D} can be rewritten as}
\begin{align}\label{MW_K}
PE[t_1, \ldots, t_N](z_1\ldots z_{{\rm rk}\,G}) = 
\prod_{I=1}^N PE\left[\chi_{R_I}(z_i) t_I\right] \,,
\end{align}
{where we used different parameters $t_1, \ldots, t_N$ for each representation $R_I$.} Here and in the following, we will omit the $B$, $F$ subscripts, as the statistics of the variables under consideration will be understood from the context. \\
Putting all together, we get
\begin{align}\label{MW_M}
P( V^G, t_1, \ldots, t_N) = \oint_{|z_i|=1} PE[t_1, \ldots, t_N](z_1\ldots z_{{\rm rk}\,G})\prod_{\vec{\alpha} \in \Delta^+} \left( 1 - \prod_{i=1}^{{\rm rk}\, G} z_i^{\alpha_i} \right)   \prod_{i=1}^{{\rm rk}\, G} \frac{dz_i}{2 \pi i z_i} \,.
\end{align}
The left-hand side is expressed as a series in the parameters $t_1, \ldots, t_N$, 
\begin{align}
\label{MW_N}
P(V^G,t_1, \ldots, t_N) = \sum_{n_I, I=1,\ldots,N} b_{n_1, \ldots, n_N} \prod_{I=1}^N t_I^{n_I} \,,
\end{align}
where the coefficients $b_{n_1, \ldots, n_N} \in \mathbb{Z}^N$ count the dimensions of each subspace at fixed power $\prod_{I=1}^N t_I^{n_I}$. The signs of the $b$'s coefficients is related to the Grassmanality of the corresponding invariant polynomial (i.e., negative signs correspond to anticommuting expressions). Notice that in general there might be no bound on the powers of the $t$'s and, being the $b$'s integer numbers, there might be cancellations among terms of fixed power $\prod_{I=1}^N t_I^{n_I}$. 

\subsection{Free Differential Algebras}

It is a well-known fact that Lie algebras can be equivalently expressed in terms of $1$-forms dual to the algebra generators $T_A$ and satisfying the MC equations. The latter can naturally be extended to include higher $p$-forms ($p >1$)
\begin{align}
\diff \sigma^i_{(p)}+\sum {\frac1n} C^i_{~i_1 \ldots i_n} 
\sigma^{i_1}_{(p_1)}\wedge \ldots \wedge \sigma^{i_n}_{(p_n)}=0,~~~~~
p+1=p_1+\ldots+p_n  \,, \label{FDA}
\end{align}
where $p, p_1,\ldots,p_n$ are the degrees of the respective forms and the indices $i, i_1,\ldots, i_n$ run on irreducible representations (irreps) of the (super)group $G$. The coefficients $C^i_{~i_1 \ldots i_n}$ are generalized structure constants satisfying generalized Jacobi identities, whose (anti)symmetry properties depend on the bosonic or fermionic nature of the forms $\sigma^{i_1},\ldots,\sigma^{i_n}$.\\
Clearly, when $p=p_1=p_2=1$ and $i, i_1, i_2$ belong to the adjoint representation of $G$, \eqref{FDA} reduces to the ordinary MC equations.\\
The equations \eqref{FDA}, together with their generalized Jacobi identities, define a FDA.
FDAs can be reformulated in terms of $L_n \subset L_\infty$-\emph{algebras}, or \emph{Strongly Homotopy Lie Algebras} {{(see \cite{Sati:2015yda,Stasheff})}}.
These generalizations of Lie algebras are proving to be essential in many contexts of physics, such as \emph{higher gauge theories} and \emph{closed String Field Theories. Their relation with FDAs was unveiled, e.g., in \cite{Stasheff,FSS} (see also \cite{Azcarraga} for a physically-oriented review).}\\
Moreover, a dual formulation of FDAs, based on a generalized Lie derivative ``along antisymmetric tensors'' has been developed in \cite{FDAdual1,FDAdual2,FDAdual3,FDAdual4} and leads to nonassociative extensions of Lie (super)algebras.

Interestingly, these generalized algebras can be obtained starting from standard Lie algebras, as shown in \cite{sullivan,DFd11,gm3,CDF}, and their construction relies on the existence of CE cohomology classes within the algebra \cite{CE}. Indeed, suppose that there exist a $p$-form
\begin{align}
\Omega^i_{~(p)} (\sigma)=\Omega^i_{~A_1 \ldots A_p} \sigma^{A_1} \ldots 
\sigma^{A_p} \,,
\end{align}
with $i$ running in an irrep of the algebra and $\Omega^i_{~A_1 \ldots A_p}$ constants. Furthermore, suppose that it is covariantly closed, but not exact, i.e.,
\begin{align}
\nabla \Omega^i_{~(p)} \equiv \diff \Omega^i_{~(p)} + \sigma^A \wedge D(T_A)^
i_{~j} \Omega^j_{~(p)}=0\,,~~~~\Omega^i_{~(p)} \not= \nabla \Phi^i_{~(p-1)
} \,.  \label{cohomology}
\end{align}
Then $\Omega^i_{~(p)}$ is said to be a representative of a Chevalley cohomology class in the $D^i_{~j}$ irrep of the algebra: here $\nabla$ is the boundary operator satisfying $\nabla^2=0$ ($\nabla^2$ is proportional to the curvature 2-form, which is zero on a Lie group).\\

The existence of $\Omega^i_{~(p)}$ allows for the extension of the original Lie algebra to the FDA defined by the following relations:
\begin{align}
&  \diff \sigma^A+\frac12 {f^A}{}_{BC} \sigma^B \wedge \sigma^C=0 \,, \nonumber\\
&   \nabla \sigma^i_{(p-1)} + \Omega^i_{(p)} (\sigma)=0 \,, \label{FDA1}
\end{align}
{where the second expression in \eqref{FDA1} should be intended as defining the new $(p-1)$-form $\sigma^i_{(p-1)}$, which cannot be expanded in the basis of the $1$-forms $\sigma^A$, but which can be added to the set of MC forms $\{\sigma^A\}$, thus enlarging the MC expression of the algebra to a FDA.}
The consistency of the above relations, namely the nilpotency of $\nabla$, is guaranteed by $\nabla \Omega^i_{(p)} = 0$.\\
Notice that if we choose a different representative of the CE cohomology class, differing from $\Omega^i_{(p)}$ by the addition of an exact term, the FDA structure remains unaltered through the redefinition $\sigma^i_{(p-1)} \rightarrow \sigma^i_{(p-1)} + \Phi^i_{(p-1)}$. \\

This procedure can be repeated on the obtained FDA, now spanned by $\sigma^A$ and $\sigma^i_{(p-1)}$, by looking for polynomials in these fields,
\begin{align}
\Omega^i_{(q)} (\sigma^A, \sigma^i_{(p-1)})= \Omega^i_{A_1 \ldots A_r i_1 \ldots i_s} \sigma
^{A_1} \wedge \ldots \wedge \sigma^{A_r} \wedge \sigma^{i_1}_{(p-1)} \wedge \ldots \wedge 
\sigma^{i_s}_{(p-1)} \,, \label{polyFDA}
\end{align}
satisfying the cohomology conditions \eqref{cohomology}. If such polynomials exists, one can extend the FDA to a new one by iterating the construction above.\\

The dynamics of the added $p$-forms is obtained, as for ordinary Lie algebras, by considering nonvanishing curvatures. For example, in this construction, $D=11$ supergravity \cite{DFd11,CDF} is based on a deformation of the FDA described by the fields $V,\omega,\psi,A{^{(3)}}$ in such a way that all curvatures are different from zero. In particular, the 3-form $A^{(3)}$ is introduced to take into account the 4-form cohomology class of $\mathfrak{siso}(1|11)/\mathfrak{so}(11)$. \\

Finally, a ``resolution'', also referred to as ``trivialization'', of a FDA can be obtained by expressing the $p$-forms, constructed in the above mentioned way, as products of $1$-forms, at the price of introducing new $1$-form fields {(without losing, however, the on-shell matching of bosonic and fermionic physical d.o.f.)}. This effectively reduces the FDA structure to a larger, but standard Lie algebra. This possibility was first considered in the seminal reference \cite{DFd11} for $D=11$ supergravity, whereas recent developments can be found in \cite{FDAnew1,FDAnew2,FDAnew3,Ravera:2021sly} (see also \cite{Penafiel:2017wfr} for related studies in lower dimensions and further developments in $D=11$).

\subsection{The work plan}\label{workplan}

In this section, we explain the interplay between the ingredients introduced above. As we have explained, a FDA is obtained from a standard Lie algebra, if one is capable of computing elements in the cohomology, thus closed but not exact cocycles. This is where the Molien-Weyl formula comes into play. A detailed explanation of the procedure is given below.\\
Given a Lie supergroup $G$, let us consider linear representations of (a subset of) its isotropy subgroup $H$ (we will always assume $H$ to be reductive in $G$) on the coset space $G/H$. The Hilbert series computes the dimensions of the space of $H'\subseteq H$-invariant polynomials, which are expressions written in terms of the 1-forms $V,\psi$, spanning the coset space $G/H$, with associated scaling weights $u,t$, respectively. It takes the following form:
\begin{align}
    P(\Omega^{\bullet}( \mathfrak{g} / \mathfrak{h}, \mathfrak{h}' ; \mathbb C ),t,u)=1+b_1t+b_2u+b_3t^2+b_4tu+b_5u^2+\ldots\,,
\end{align}
where the coefficients $b_1,b_2,\ldots$ in the above expression indicate the number of independent invariant polynomials constructed with the $1$-forms at our disposal (the sign included in such coefficients specify the commutation ($+$) or anticommutation ($-$) property of the invariants). The explicit form of such polynomials in terms of $V,\psi$ can be inferred by inspecting the corresponding powers in $u,t$.\\
As the building 1-forms $V,\psi$ are constrained to satisfy the MC equations, the corresponding scaling weights are not unrelated. In general, if $\alpha$ is a generic index in a representation of $H$, under which the gravitini transform, the corresponding MC equations read
\begin{align}\label{FF_C}
\nabla V^a &\equiv \diff V^a + \omega^a_{~b} V^b =f^a_{\alpha \beta} \psi^\alpha \psi^\beta\,, \nonumber\\
\nabla \psi^\alpha &\equiv \diff \psi^\alpha + \omega^\alpha_{~\beta} \psi^{\beta}  =   f^\alpha_{b \beta} V^b \psi^\beta \,,
\end{align}
where $\omega^a_{~b}$ and $\omega^\alpha_{~\beta}$ are the spin connections for the vector and spinor representations, respectively. \\
Let us now distinguish two possible cases.

{\flushleft\bf Flat rigid superspaces: $f^\alpha_{b \beta}=0$.}\\
A convenient choice of relation between the weights $u$ of the vielbein and $t$ of the gravitino, compatible with the MC equations and avoiding attributing a scale to the differential $\diff$, is $u=t^2$.
If 
\begin{align}
    P(\Omega^{\bullet}( \mathfrak{g}/\mathfrak{h} , \mathfrak{h'} ; \mathbb C ),t,u(t))=\lim_{u\to u(t)}P(\Omega^{\bullet}( \mathfrak{g}/\mathfrak{h} , \mathfrak{h'} ; \mathbb C ),t,u)=\sum_nb_nt^n\,,
\end{align} 
that is the Hilbert-Poincaré series evaluated with $u=u(t)$ ab initio coincides with the one where $u=u(t)$ is plugged in only after the Molien-Weyl integration, then the series automatically selects, out of all the possible invariant polynomials, only those which correspond to cohomology classes and, therefore, to nonexact cocycles.\\
In order to understand why this is the case, we have to rewrite the above expression in a different manner, that is\footnote{The present exposition is adapted from \cite{Berkovits:2005hy}.} 
\begin{align}
\label{cippaA}
   P(\Omega^{\bullet}( \mathfrak{g}/\mathfrak{h} , \mathfrak{h'} ; \mathbb C ),t,u(t)) = {\rm Tr}_{H}[ (-1)^F t^K \rho({g})] \,,
\end{align}
where $H(\Omega^{\bullet}( \mathfrak{g} , \mathfrak{h} ; \mathbb C))$ is the CE cohomology, $K$ is 
the dimension operator that counts the dimension of the MC forms (e.g., $K[\psi^\alpha]=1$), and $F$ is the fermionic charge operator. The formula computes the trace of the group element in the suitable 
representation (see the previous discussion for the explicit computation of the trace). Now, since both $K$ and $F$ define a grading and since the differential must respect it, the cohomology $H$ is filtered according to those gradings, $H = \oplus_{n,f} H^{n,f}$, where $H^{n,f}$ denote the cohomology groups with  fermionic number $f$ and dimension $n$. Then, 
\eqref{cippaA}
becomes 
\begin{align}\label{cippaB}
  P(\mathfrak{g}/\mathfrak{h} , \mathfrak{h'} ; \mathbb C ),t,u(t))  = \sum_{n} t^n  \sum_{f=0,1} (-1)^f 
   {\rm Tr}_{H^{n,f}}[\rho({g})] 
\end{align}
and therefore the numbers $b_n$ are computed by $\sum_{f=0,1} (-1)^f 
   {\rm Tr}_{H^{n,f}}[\rho({g})]$. 
The claim is that 
\begin{align}\label{cippaC}
b_n = \sum_{f=0,1} (-1)^f  {\rm Tr}_{\mathcal{F}^{n,f}}[\rho({g})] \,,
\end{align}
where $\mathcal{F}^{n,f}$ are all elements of the space $\Omega^{\bullet}( \mathfrak{g} , \mathfrak{h} ; \mathbb C)$ which are not 
necessarily in the cohomology, filtered with respect to the dimensions and fermionic number. 
This means that the computation of $b_n$ can be done by summing over all possible 
representatives. To prove this claim, observe that
\begin{align}\label{cippaD}
{\rm Tr}_{\mathcal{F}^{n,f}} = {\rm Tr}_{\mathcal{Z}^{n,f}} + {\rm Tr}_{\mathcal{F}^{n,f}/\mathcal{Z}^{n,f}} = 
 {\rm Tr}_{\mathcal{H}^{n,f}} + {\rm Tr}_{\mathcal{B}^{n,f}} +  {\rm Tr}_{\mathcal{B}^{n,f+1}} \,,
\end{align}
where $\mathcal{Z} = {\rm Ker} \, \diff$ and $\mathcal{B} ={\rm Im} \, \diff$ and, therefore, the last two addenda of \eqref{cippaD} 
drop out from the alternating sum \eqref{cippaC}. This implies that the integration performed in the Molien-Weyl formula is over the space $\mathcal{F}$ and this is enough to select the cohomology. 

{\flushleft\bf Curved rigid superspaces: $f^\alpha_{b \beta}\neq0$}\\
In this case, the presence of a cosmological constant in the associated supergravity theory forces the attribution of a scale to the differential $\nabla$, or to the structure constants. {In the following, we will opt for the former case, as it will ease the interpretation of the resulting Hilbert series in terms of form numbers.}
A consistent choice of the scaling weights is the following: $u=t$ and $[\nabla]=t$. Notice, in particular, that, due to the latter constraint, different invariant polynomials corresponding to different powers in the Hilbert-Poincaré series, might be related by $\nabla.$\\
Despite this, the Molien-Weyl formula greatly simplifies the quest of finding elements of the cohomology group, as it allows to restrict the number of possible invariant polynomials to check. \\

Independently of the structure of the rigid superspaces, to any class $\omega^{(p)}$ computed in the above mentioned ways, one can associate a $(p-1)$-form $A^{(p-1)}$, whose scaling can be inferred from the one of $\omega^{(p)}$. Depending on the commutation/anticommutation property of $A^{(p-1)}$, one divides/multiplies the Hilbert series {by an appropriate plethystic polynomial, accounting for the addition of the new field. This procedure yields a new series,} describing new cocycles and written in terms of $V$, $\psi$, $A^{(p-1)}$.
This procedure can go on iteratively and there are two possible outcomes: either the Hilbert series trivializes, namely it becomes $1$, after a finite number of steps, or one is forced to introduce an infinite number of additional forms to completely trivialize it.\\

We finally remark that in the following we will work in Euclidean signature. This is motivated by the fact that determining roots and weights of the various representations will require us to consider algebras (and cosets obtained by quotients of algebras) over the complex field $\mathbb{C}$. 

\section{$D=4$ spacetime dimensions}\label{sec3}

In this section, we will focus on the study of the cohomology groups for superspaces with four bosonic dimensions, with various amounts of supersymmetry. We will first discuss flat cases and only then describe our results of curved supermanifolds.

\subsection{Flat rigid superspaces}

As discussed in Sec. \ref{workplan}, the Molien-Weyl formula for flat rigid superspaces automatically selects all the cohomologies out of the invariant polynomials computed by the Hilbert-Poincaré series.

\subsubsection{$N=1$ supersymmetry}\label{D4N1flat}

Let us start this analysis by considering the simplest case: $D=4$, $N=1$ flat superspace, corresponding to the coset space $G/H=sISO(1|4)/SO(4)$. The algebra associated with $sISO(1|4)$ is the super-Poincaré one in $D=4$, which contains 4 translation generators, 4 supersymmetry generators, and the Lorentz subalgebra.\\
The coset can be dually described in terms of the supervielbein $(V_{\alpha \dot \alpha}, \psi_{\alpha}, \bar\psi_{\dot \alpha})$, satisfying
\begin{align}\label{N1A}
\diff V_{\alpha \dot \alpha} =  
\psi_{\alpha} \bar\psi_{\dot \alpha}\,,
~~~~~
\diff \psi_{\alpha} =0\,, ~~~~~
\diff \bar\psi_{\dot \alpha} =0\,,
\end{align}
where $\alpha,\dot\alpha$ are indices transforming in the spinorial representation of each of the two $SL(2,\mathbb C)$ factors in $SO(4)$, whereas $(\alpha\dot\alpha)$ is a vector index (remember that in the following we will not distinguish between $SU(2)$ and $SL(2,\mathbb C)$, as specified above).  
In this notation, $\psi_\alpha$ is associated with the chiral component of the gravitino and $\bar\psi_{\dot\alpha}$ with its antichiral complex conjugate, while
$V^{\alpha\dot\alpha}$ is defined in terms of the usual bosonic vielbein $V^a$ as {$V^{\alpha\dot\alpha}= -\frac i2\bar\gamma^{\dot\alpha\alpha}_a V^a$}, $\gamma^{\alpha\dot\alpha}_a$ being the Pauli matrices.\footnote{In Lorentzian signature, one usually defines $\gamma^a_{\alpha\dot\alpha}=(\mathbb 1,\gamma^i)$ and $\bar\gamma^{a\dot\alpha\alpha}=\epsilon^{\alpha\beta}\epsilon^{\dot\alpha\dot\beta}\gamma^a_{\beta\dot\beta}=(\mathbb 1,-\gamma^i)$, with $\epsilon^{12}=\epsilon_{12}=1$. The Euclidean case is easily obtained by considering $\gamma^i\to i\gamma^i$ and the new matrices satisfy the following completeness relation:
$$\gamma^a_{\alpha\dot\alpha}\gamma^b_{\beta\dot\beta}\delta_{ab}=2\epsilon_{\alpha\beta}\epsilon_{\dot\alpha\dot\beta}\,.$$}\\
{Moreover, notice that we are using the differential $\diff$, as the spin connection vanishes for flat spacetimes.}\\

Since we want to consider invariant polynomials written in terms of the supervielbein spanning the coset space $G/H$, we need to start from the characters associated with the vielbein and gravitini. These can be computed as explained in Sec. \ref{sec2} and read
\begin{align}\label{N1B}
 \chi_V(z,w) = \left( z + \frac1z\right) \left(w + \frac1w\right) \,, \qquad \chi_\psi(z)= 
 \left(z + \frac1z\right) \,, \qquad \chi_{\bar\psi}( w) =
 \left(w + \frac1w\right) \,,
\end{align}
whose corresponding plethystic polynomials are
\begin{align}
\label{pleA}
PE[\chi_V u] &= (1 - u z) \left(1 - \frac{u}{z} \right) (1 - u w)  \left(1 - \frac{u}{w} \right)\,, \nonumber \\
{PE}[\chi_\psi t] &= \left[(1 - t z) \left(1 - \frac{t}{z} \right)\right]^{-1} \,, \qquad {PE}[\chi_{\bar\psi} t] = \left[(1 - t w)  \left(1 - \frac{t}{w} \right)\right]^{-1}\,.
\end{align}
The other needed ingredient for the computation of the Molien-Weyl formula is the integration measure. Depending on the choice of the latter, one can in general select invariants with respect to different subsets of the subgroup $H$. We are here interested in invariants under the full Lorentz group $SO(4)$: the invariant measure associated with a single $SU(2)$ factor is given by 
\begin{align}
 \label{newE}
 \diff \mu_{SU(2)} = (1 - w^2)\frac{dw}{(2 \pi i) w}\,,
\end{align}
while the full measure is easily obtained as 
\begin{align}
\label{newUAA1}
\diff \mu_{SU(2)\times SU(2)} = (1 - w^2)(1 - z^2) \frac{dw}{(2 \pi i) w} \frac{dz}{(2 \pi i) z}\,.
\end{align}
The Hilbert-Poincaré polynomial for the case $\mathfrak{g}/\mathfrak{h}=\mathfrak{g}/\mathfrak{h}'=\mathfrak{siso}(1|4)/ \mathfrak{so}(4)$ is then given by
\begin{align}\label{pleB}\nonumber
P(\Omega^{\bullet}(\mathfrak{g}/\mathfrak{h}, \mathfrak{h}' ; \mathbb C ),t,u) &= \oint_{|z|=1} \oint_{|w|=1} PE[\chi_V u]{PE}[\chi_\psi t] 
{PE}[\chi_{\bar\psi} t] \diff \mu_{SU(2)\times SU(2)} \,\\
&= 1 - t^2 (1-u)^2 u + u^4\,.
\end{align}
As explained in Sec. \ref{workplan}, each monomial indicates the weight of the invariant polynomials belonging to the {respective} vector subspace, whereas the coefficient is the dimension of the invariant vector subspace and tells how many independent polynomials there are. Finally, the sign corresponds to their parity, i.e., their commuting/anticommuting'' properties. We list below all the possible independent invariant polynomials corresponding to such polynomial:
\begin{align}\nonumber\label{N1CA}
    - t^2 u: \qquad \omega^{(3)} &= V^{\alpha\dot\alpha} \psi_{\alpha} \bar \psi_{\dot\alpha}\,, \\ \nonumber
    2 t^2 u^2: \qquad \omega^{(4)}_1 &= V^{\alpha\beta}_2 \psi_{\alpha} \psi_{\beta}\,, ~~~~~~
\omega^{(4)}_2 = V_2^{\dot\alpha\dot\beta} \bar\psi_{\dot\alpha} \bar\psi_{\dot\beta}\,, \\ \nonumber
    - t^2 u^3: \qquad \omega^{(5)} &= V_3^{\alpha\dot\alpha} \psi_{\alpha} \bar\psi_{\dot\alpha}\,,\\
    u^4: \qquad \omega^{(4)} &= V_4\,,
\end{align}
where we defined $V_2^{\alpha\beta} = \frac12 \epsilon_{\dot\alpha\dot\beta}V^{\alpha\dot\alpha} \wedge V^{\beta\dot\beta}$, 
$V_2^{\dot\alpha\dot\beta} = \frac12 \epsilon_{\alpha\beta}V^{\alpha\dot\alpha} \wedge V^{\beta\dot\beta}$, $V_3^{\alpha\dot\alpha} = \frac{1}{3!} (V \wedge V \wedge V)^{\alpha\dot\alpha}$, and $V_4 = \frac{1}{4!} V^{\alpha \dot{\alpha}} \epsilon_{\alpha \beta} V^{\beta \dot{\beta}} \epsilon_{\dot{\beta} \dot{\gamma}} V^{\gamma \dot{\gamma}} \epsilon_{\gamma \delta} V^{\delta \dot{\delta}} \epsilon_{\dot{\alpha} \dot{\delta}}$. \\
The fact that the vielbein and gravitino satisfy the MC equations \eqref{N1A} requires a relation between the scaling weights $u$ and $t$ introduced by the plethystic polynomials. In particular, as argued above, we consider the compatibility choice $u(t)=t^2$. Hence, the Hilbert-Poincaré polynomial \eqref{pleB} boils down to
\begin{align}\label{N1C}
 P(\Omega^{\bullet}(\mathfrak{g}/\mathfrak{h} , \mathfrak{h'} ; \mathbb C ),t,t^2) =1 - t^4 + 2 t^6 \,.
\end{align}
The resulting series indicates the presence of three cohomology classes out of the invariant polynomials in \eqref{N1CA}: these are $\omega^{(3)}$, $\omega^{(4)}_1$, $\omega^{(4)}_2$, as one can prove that 
\begin{eqnarray}\label{N1D}
\diff \omega^{(4)} = \omega^{(5)}\,, ~~~~ \diff \omega^{(5)} = 0\,.
\end{eqnarray}
As explained in Sec. \ref{workplan}, one now has to add to the complex of differential forms some new{ potentials, which, through their exterior derivatives, trivialize the corresponding cohomology.} In particular, we can add an even potential $B^{(2)}$ and two odd potentials $A^{(3)}_i$ satisfying
\begin{eqnarray}
 \label{N1E}
 \diff B^{(2)} = \omega^{(3)}\,, ~~~~~
 \diff A^{(3)}_i = \omega^{(4)}_i\,, ~~~i=1,2 \,,
\end{eqnarray}
where $B^{(2)}$ and $A^{(3)}_i$ carry dimensions $t^4$ and $t^6$, respectively. These new generators enlarge the (bosonic) dimension of the superspace we are representing $SO(4)$ on. \\
{The physical interpretation of these objects can be related to the construction of Wess-Zumino terms for the Green-Schwarz superstring in $D=4$, see \cite{AETW,FSS,DUFF1,DUFF2,Bandos:2019wgy}. \\
In particular, one can define the following Wess-Zumino couplings
\begin{align}\label{more4C}
\Gamma_{WZ} \sim \int_{\mathcal {M}^{(3)}} \omega^{(3)}=\int_{\partial\mathcal M^{(3)}}B^{(2)},\qquad \Gamma_{WZ,i} \sim \int_{\mathcal {M}^{(4)}} \omega^{(4)}_i=\int_{\partial\mathcal M^{(4)}}A^{(3)}_i\,,
\end{align}
where $\mathcal M^{(3)}$, $\mathcal M^{(4)}$ are manifolds with a boundary.}\\
The new series describing the number of invariants written in terms of $V$, $\psi$, $\bar\psi$, $B^{(2)}$, and $A^{3}_i$ is
\begin{equation}\label{N1CB}
P_{{\rm FDA}}(\Omega^{\bullet}(\mathfrak{g}/\mathfrak{h} , \mathfrak{h'} ; \mathbb C ),t,{t^2}) = 
 (1 - t^4 + 2 t^6) \frac{(1-t^6)^2}{(1-t^4)} \,,
\end{equation}
where the factor $\frac{1}{(1-t^4)}$ corresponds to the contribution of the commuting potential $B^{(2)}$ and $(1-t^6)^2$ comes from the anticommuting $A^{(3)}_i$ forms.\\
The denominator can be expanded as a geometric series and takes into account all possible powers of $B^{(2)}$, while $(1-t^6)^2$ determines the powers of $A^{(3)}_i$, $A^{(3)}_1\wedge A^{(3)}_2$. However, by expanding the new Hilbert series in terms of $t$, we get, for the few first terms, 
\begin{equation}\label{N1CBA}
P_{{\rm FDA}}(\Omega^{\bullet}(\mathfrak{g}/\mathfrak{h} , \mathfrak{h'} ; \mathbb C ),t,{t^2}) = 1+2 t^{10}-3 t^{12}+2 t^{14}-4 t^{16}+4 t^{18}-4 t^{20}+O\left(t^{21}\right) \,,
\end{equation}
showing that there are new cohomology classes created by the new potentials. \\
To proceed with the FDA construction, we should now add other potentials to compensate for the (infinite number of) cohomology classes.
The above computation shows that, in fact, one needs an infinite number of {super}forms to trivialize the FDA series. The dimension of each space at fixed power of $t$ can be obtained by means of the Poincaré-Birkhoff-Witt theorem \cite{Connes:2002ya,Berkovits:2005hy}, which trivializes the polynomial in the following way:
\begin{equation}
    P_{{\rm FDA}}(\Omega^{\bullet}(\mathfrak{g}/\mathfrak{h} , \mathfrak{h'} ; \mathbb C ),t,{t^2})=\frac{P(\Omega^{\bullet}(\mathfrak{g}/\mathfrak{h} , \mathfrak{h'} ; \mathbb C ),t{,t^2})}{\prod_{p\geq 1}(1- t^{2p})^{N(p)}}=1 \,,
\end{equation}
where $N(p)$ is given by 
\begin{eqnarray}\label{N1F}
 N(p) = \sum_{n|p} \mu\left(\frac{p}{n}\right) (s_1^n + s_2^n + s_3^n)  
\end{eqnarray}
and $n|p$ are the divisors of $p$ and the sum is extended to the entire set of the divisors of each $p$, $\mu(x)$ is the M\"obius function and $u_i$ are the three roots of the polynomial $(1 - s/2 + s^3/2)$ (which is 
related to the original equation as $(1 - t^4 + 2 t^6) = 2 t^6 (1 - s/2 + s^3/2)$ with $s = t^{-2}$). With a simple computation, one gets the first terms of $N(p)$ as 
\begin{eqnarray}\label{N1G}
 N(p) &=& \left(0,1,-2,0,-2,3,-2,4,-4,7,-10,11,-18,27,-34,50,-74,\right. \nonumber \\
&&\left.106,-154,217,-318,471,-674,978,\right. \nonumber \\
&&\left.-1440,2111,-3084,4511,-6642,9791, \ldots
 \right) \,.
\end{eqnarray}
As an example of the above formula, we see that the first cohomology classes require the introduction of one commuting potential $B^{(2)}$ with dimension $t^4$ and two anticommuting potentials $A^{(3)}_i$ with dimension $t^6$, matching the above results.
The next necessary forms to be added are two anticommuting potential $C^{(5)}$ with dimension $t^{10}$ and so on.\\
Let us build explicitly the representatives for the first classes displayed in \eqref{N1CBA}. 
We start from $B^{(2)}$, $A^{(3)}_i$, $V^a$, $\psi$, $\bar\psi$ and the first terms in the series are the following cocycles: 
\begin{align}\label{more4A}
2 t^{10}: \quad & {\omega^{(6)}_1 = \omega^{(3)}\wedge A_1^{(3)} \,, ~~~~~
\omega^{(6)}_2 = \omega^{(3)}\wedge A_2^{(3)}} \,, \nonumber \\
- 3 t^{12}: \quad & \omega^{(7)}_1 = \omega^{(4)}_1\wedge A_1^{(3)}\,, ~~~~
  \omega^{(7)}_2 =\omega^{(4)}_2\wedge A_2^{(3)}\,, ~~~~
 \omega^{(7)}_3 = \omega^{(4)}_2 \wedge A_1^{(3)} + \omega^{(4)}_1\wedge A_2^{(3)}\,,
 \nonumber \\
2 t^{14}: \quad  
& \omega^{(8)}_1 = \omega^{(4)}_1\wedge B^{(2)} \wedge B^{(2)}\,, ~~~~
  \omega^{(8)}_2 =\omega^{(4)}_2\wedge B^{(2)} \wedge B^{(2)}\,. 
\end{align}
The powers of $B^{(2)}$'s are coming from the fact that $B^{(2)}$ is a commuting 2-form and therefore, in principle, we can consider higher powers of them.\\
It is worth mentioning that forms with degree higher than 4 are \emph{pure soul} \cite{Deligne}, i.e., their pull-back on spacetime is zero independently of their embedding in superspace. These higher form cocycles are perfectly well-defined from an algebraic point of view, but their physical interpretation is yet to be fully understood.
Nonetheless, a possible application can be suggested in terms of nonfactorized actions \cite{CG1,CG2}, that is functionals that cannot be expressed in terms of the pull-back of Lagrangian densities. For example, one could consider $\omega^{(p)}\wedge\star\,\omega^{(p)}$, which makes sense $\forall p\in \mathbb N$, once integrated on a given metric-supermanifold \cite{CCG}. 

\subsubsection{$N=2$ supersymmetry}

Let us now increase the number of supersymmetry generators and let us consider the coset superspace ${(U(2)\ltimes sISO(2|4))/(SO(4)\times U(2))}$, spanned by the supervielbein $V_{\alpha \dot \alpha}, \psi_{\alpha}^A, \bar\psi_{\dot \alpha, A}$.\footnote{The action of the $R$-symmetry group realises only on the even forms as shown, e.g., in \cite{Frappat}.} The following MC equations hold
\begin{eqnarray}\label{N2A}
\nabla V_{\alpha \dot \alpha} =  
\psi_{\alpha}^A \bar\psi_{\dot \alpha A}\,,
~~~~~
\nabla \psi_{\alpha}^A =0\,, ~~~~~
\nabla \bar\psi_{\dot \alpha A} =0\,. 
\end{eqnarray}
where $A=1,2$ is a $U(2)$ $R$-symmetry index transforming in the $[0,1/2]$ and $[1/2,0]$ representations for the gravitini $\psi_{\alpha}^A$ and $\bar\psi_{\dot \alpha A}$ respectively. Let us notice that, due to the absence of a $U(2)$-invariant tensor, the position of the indices for the chiral gravitini is fixed.
As in the $N=1$ case, the $SO(4)$ indices can be split in terms of $SU(2)$ indices.\\
The covariant derivative is nilpotent $\nabla^2=0$, despite containing the gauge connection associated with the $U(2)$ factor. In fact, the square of the covariant derivative would in general close on the curvature of the subalgebra we are modding out, which however vanishes in the present case. \\
The characters of the supervielbein are 
\begin{align}\label{N2B}\nonumber
 \chi_V(z,w) &= \left( z + \frac1z\right) \left(w + \frac1w\right) \,, \\ \nonumber
 \chi_\psi(z_1,z_2, z) &= z_1 \left(z_2 + \frac{1}{z_2}\right)
 \left(z + \frac1z\right) \,, \\
 \chi_{\bar\psi}(z_1, z_2, w) &= \frac{1}{z_1}\left(z_2 + \frac{1}{z_2}\right)
 \left(w + \frac1w\right) \,,
\end{align}
where $z,w$ parametrize the Lorentz group, $z_2$ is related to $SU(2)\subset U(2)$ and $z_1$ parametrizes the additional $U(1)$ $R$-symmetry factor. Let us now consider different sets of invariant polynomials.\\
$\bullet$ Taking $\mathfrak{g}/\mathfrak{h}=(\mathfrak{u}(2)\ltimes \mathfrak{siso}(2|4))/(\mathfrak{so}(4)\oplus\mathfrak{u}(2))$ and $\mathfrak{h}'=\mathfrak{so}(4)\oplus\mathfrak{u}(1)$, we get the following Hilbert-Poincaré polynomial:
\begin{align}\label{RinvA}
P(\Omega^\bullet(\mathfrak{g}/\mathfrak{h},\mathfrak{h}';\mathbb C),t,t^2,z_2) = \frac{1+t^2}{1-t^2}   - t^4 (z_2^2 + 1 + \frac{1}{z_2^2}) \,.
\end{align}
The first term in the above expression reproduces the $SU(2)$-invariant polynomials, which will be carefully analyzed below, whereas the second term corresponds to a Lorentz-invariant polynomial in the $SU(2)$ triplet representation, whose form is
\begin{align}\label{RinvB}
\omega^{(3)A}{}_B = \bar\psi_{\dot \alpha B} V^{\alpha\dot \alpha } \psi^A_{\alpha} -  
\frac12 \delta^A_B \, 
\bar\psi_{\dot \alpha C} V^{\dot \alpha \alpha} \psi^{C}_{\alpha} \,.
\end{align}
Notice that $\omega^{(3)1}{}_1=-\omega^{(3)2}{}_2$. Furthermore, from Sec. \ref{workplan}, we know that these forms actually correspond to cohomology classes once the MC equation hold, as it can be checked in this case. One can then introduce three even $2$-forms satisfying 
\begin{align}
    \diff A^{(2)A}{}_B+\omega^{(3)A}{}_B=0 \,.
\end{align}
The FDA polynomial obtained in this way is actually a series, analogously to what we have obtained in the $N=1$ case. \\
$\bullet$ Let us now consider $\mathfrak{h}'=\mathfrak{so}(4)\oplus\mathfrak{su}(2)$. In this case, the Hilbert-Poincaré series reads
\begin{align}\label{N2C0}
 P(\Omega^\bullet(\mathfrak{g}/\mathfrak{h},\mathfrak{h}';\mathbb C),t,u,z_1)  = \frac{1 - t^2 (u + u^3) + u^4}{(1-t^2)^2} \,,
\end{align}
{where the dependence on $z_1$ automatically disappears.} The above series has the following generators:
\begin{align}\label{N2D}
    {\frac{1}{(1-t^2)^2}}:\qquad \omega^{(2)}_1&=  \epsilon_{AB}\epsilon^{\alpha\beta}\psi_\alpha^A \psi_\beta^B \,, \qquad \omega^{(2)}_2= \epsilon^{AB}\epsilon^{\dot\alpha\dot\beta}\bar \psi_{\dot\alpha  A}\bar\psi_{\dot\beta B} \,, ~~~~~\nonumber \\
    - t^2 u:\qquad \omega^{(3)}&= V^{\alpha\dot\alpha} \psi_{\alpha}^A \bar \psi_{\dot\alpha A}\,, \nonumber \\
    - t^2 u^3:\qquad \omega^{(5)}&= V_3^{\alpha\dot\alpha} \psi_{\alpha}^A \bar \psi_{\dot\alpha A}\,, \nonumber \\
    u^4: \qquad \omega^{(4)}&= V_4\,,
\end{align}
where $\epsilon_{AB}$ is the $SU(2)$-invariant tensor. When the MC are satisfied, one can choose, as explained, $u(t)=t^2$ and the series reduces to
\begin{eqnarray}\label{N2C}
 P(\Omega^\bullet(\mathfrak{g}/\mathfrak{h},\mathfrak{h}';\mathbb C),t,t^2,z_1) = \frac{1+t^2}{1-t^2} = 1 + \frac{2{t^2}}{1-t^2} \,,
\end{eqnarray}
as anticipated in the previous point. The MC equations select, out of all the invariant polynomials, only the cohomologies: one can indeed prove that
\begin{align}\label{N2D1}
\diff  \omega^{(2)}_1 &= 0\,, \quad \diff  \omega^{(2)}_2 = 0\,,\quad \diff  \omega^{(3)} = \omega^{(2)}_1 \wedge \omega^{(2)}_2\,, \quad \diff \omega^{(4)} = \omega^{(5)} \,.
\end{align}
The associated FDA is obtained by including two independent $1$-forms, $A^{(1)}_1$ and $A^{(1)}_2$, such that 
\begin{align}\label{N2D2}
\diff A^{(1)}_1 = \omega^{(2)}_1\,, \qquad
\diff A^{(1)}_2 = \omega^{(2)}_2\,, 
\end{align}
leading to the following series:
\begin{eqnarray}\label{N2E}
 P_{\rm FDA}(\Omega^\bullet(\mathfrak{g}/\mathfrak{h},\mathfrak{h}';\mathbb C),t,t^2,t^0) =   \frac{1+t^2}{1-t^2}  (1-t^2)^2 = 1-t^4 \,.
\end{eqnarray}
This corresponds to the following cohomology class:
\begin{align}
\widetilde\omega^{(3)} = A^{(1)}_1 \wedge \omega^{(2)}_2 +  A^{(1)}_2 \wedge \omega^{(2)}_1 - 2  \omega^{(3)}\,.
\end{align}
The addition of a commuting $2$-form $B^{(2)}$ completely trivialises the Hilbert series:
\begin{eqnarray}\label{N22}
 P'_{\rm FDA}(\Omega^\bullet(\mathfrak{g}/\mathfrak{h},\mathfrak{h}';\mathbb C),t,t^2,z_1) =   \frac{1-t^4}{1-t^4}  =1 \,.
\end{eqnarray} \\
$\bullet$ At last, let us consider $\mathfrak{h}'=\mathfrak{so}(4)\oplus\mathfrak{u}(2)$. The Hilbert-Poincaré series in this case is 
\begin{align}
 \label{N2F0}
 P(\Omega^\bullet(\mathfrak{g}/\mathfrak{h},\mathfrak{h}';\mathbb C),t,u)=\frac{1 - t^2 (u + u^3) + u^4}{(1-t^4)}\,,
\end{align}
where the only change, compared with the previous case, is in the denominator. When the MC equations are implemented we get
\begin{align}
 \label{N2F}
 P(\Omega^\bullet(\mathfrak{g}/\mathfrak{h},\mathfrak{h}';\mathbb C),t,t^2)=  1\,.
\end{align}
Therefore, there are no invariant cocycles in the cohomology in the present case.

\subsubsection{$N=4$ supersymmetry}

Let us now further increase the amount of supersymmetry to $N=4$, due to the important relation between $D=6$, $N=(2,0)$ and $D=4$, $N=4$ super Yang-Mills (see, e.g., the original paper \cite{Sohnius:1981sn}). {In the latter theory, the reality condition on the scalar fields of the vector supermultiplet, necessary for the matching of bosonic and fermionic degrees of freedom, reduces the $R$-symmetry group from $U(4)$ to $SU(4)$. For this reason, let us consider here} a supervielbein spanning the cotangent space of the coset ${(SU(4)\ltimes sISO({4}|4))/(SO(4)\times SU(4))}$ satisfying
\begin{eqnarray}\label{N4A}
\nabla V_{\alpha \dot \alpha} =  
\psi_{\alpha}^A \bar\psi_{\dot \alpha A}\,,\qquad \nabla \psi_{\alpha}^A =0\,, \qquad \nabla \bar\psi_{\dot \alpha A} =0\,. 
\end{eqnarray}
The covariant differential behaves as in the $N=2$ case and the index $A=1,\ldots,4$ of $\psi_{\alpha}^A$ and $\bar\psi_{\dot \alpha A} $ transforms in the $[0,0,1]$ and $[1,0,0]$ representations. The characters associated with the supervielbein are 
\begin{align}\label{N4B}
 \chi_V(z,w) &= \left( z + \frac1z\right) \left(w + \frac1w\right) \,, \nonumber \\
 \chi_\psi(z_1, z_2, z_3, z) &= \left(z_1 + \frac{z_2}{z_1} + \frac{z_3}{z_2} + \frac{1}{z_3}\right)
 \left(z + \frac1z\right) \,, \nonumber \\
  \chi_{\bar\psi}(z_1, z_2, z_3, w) &= \left(z_3 + \frac{z_2}{z_3} + \frac{z_1}{z_2} + \frac{1}{z_1}\right)
 \left(w + \frac1w\right) \,,
\end{align}
where $z$ and $w$ parametrize the two $SU(2)$ factors in the Lorentz group, whereas $z_1$, $z_2$, and $z_3$ refer to the $SU(4)$ group.
The measure for the latter is 
\begin{align}\label{newE1}
 \diff \mu_{SU(4)} = \left(1-\frac{z_1^2}{z_2}\right) \left(1-\frac{z_1 z_2}{z_3}\right) \left(1-z_1 z_3\right) \left(1-\frac{z_2^2}{z_1 z_3}\right) \left(1-\frac{z_2 z_3}{z_1}\right) \left(1-\frac{z_3^2}{z_2}\right) \frac{dz_1dz_2 dz_3}{(2 \pi i)^3 z_1 z_2 z_3} \,.
\end{align}
As we have done in the $N=2$ case, we will consider different sets of invariant polynomials.\\
$\bullet$ The cohomology invariants under the Lorentz group are found by inspecting the following Hilbert series (where $\mathfrak{g}/\mathfrak{h}=(\mathfrak{su}(4)\ltimes \mathfrak{siso}(4|4))/(\mathfrak{so}(4)\oplus \mathfrak{su}(4))$ and $\mathfrak{h}'=\mathfrak{so}(4)$):
\begin{align}\label{N4AA}
&P(\Omega^\bullet(\mathfrak{g}/\mathfrak{h},\mathfrak{h}';\mathbb C),t,t^2,z_1, z_2, z_3)=
1+2 \left( \frac{ z_1}{z_3}+\frac{ z_3 z_1}{z_2}+ z_2+\frac{1}{z_2}+\frac{z_2}{z_1 z_3}+\frac{z_3}{z_1} \right)   t^2 \nonumber \\ 
&+\frac{3}{z_1^2 z_2^2 z_3^2}{\left(z_3^4 z_1^4+z_2^2 z_1^4+z_2 z_3^2 z_1^4+z_2^2 z_3^3 z_1^3+z_3^3 z_1^3+z_2^3 z_3 z_1^3+z_2 z_3 z_1^3+ z_2 z_3^4 z_1^2 +z_2^3 z_1^2\right.} \nonumber \\
&{ \left. 
+z_2^4 z_3^2 z_1^2+2 z_2^2 z_3^2 z_1^2+z_3^2 z_1^2+z_2^3 z_3^3 z_1+z_2 z_3^3 z_1+z_2^4 z_3 z_1+z_2^2 z_3 z_1+z_2^4+z_2^2 z_3^4+z_2^3 z_3^2\right)} t^4  +O\left(t^5\right)
\end{align}
at order $t^4$. The first term corresponds to cohomologies which are invariant under $SO(4)\times SU(4)$, which will be analyzed below, whereas all other terms are only invariant under the Lorentz group. For example, the second term corresponds to two independent cohomologies transforming in the $4$ and $\bar 4$-antisymmetric representations, respectively, and taking the following form:
\begin{align}\label{N4AB}
\omega^{(2) [AB]}_{1} = \epsilon^{\alpha\beta} \psi^A_\alpha \psi^B_\beta\,, \qquad
\omega^{(2) }_{2}{}_{[AB]} = \epsilon^{\dot\alpha\dot\beta} \bar\psi_{A \dot\alpha}\bar \psi_{B\dot\beta}\,. ~~~
\end{align}
These objects can be trivialized by introducing two sets of corresponding $1$-form, 
\begin{eqnarray}\label{N@A}
\diff A^{(1) [AB]} = \omega^{(2) [AB]}_1 \,, ~~~~~~ \diff A^{(1)}{}_{[AB]} = \omega^{(2)}_2{}_{[AB]}\,, 
\end{eqnarray}
which can be understood as the $N=4$ gauge $1$-forms in the gravitational supermultiplet. \\
$\bullet$ The integration over the full subgroup $SO(4)\times SU(4)$ yields the following series (we have $\mathfrak{h}'=\mathfrak{so}(4)\oplus\mathfrak{su}(4)$):
\begin{eqnarray}\label{N4C0}
 P(\Omega^\bullet(\mathfrak{g}/\mathfrak{h},\mathfrak{h}';\mathbb C),t,u) = \frac{1 - t^2 (u + u^3) + u^4}{(1-t^4)}\,,
\end{eqnarray}
corresponding to the invariant expressions
\begin{align}\label{newOF}
\frac{1}{1-t^4}:\qquad \omega^{(4)}_1 &= 
\left( \epsilon^{\alpha\beta}\psi_\alpha^A \psi_\beta^B\right) 
\left( \epsilon^{\dot\alpha\dot\beta}\bar \psi_{\dot\alpha A} 
\bar\psi_{\dot \beta B}\right) 
\,, ~~~~~\nonumber \\
- t^2 u:\qquad \omega^{(3)} &= V^{\alpha\dot\alpha} \psi_{\alpha}^A \bar \psi_{\dot\alpha A}\,, \nonumber \\ 
- t^2 u^3: \qquad \omega^{(5)} &= V_3^{\alpha\dot\alpha} \psi_{\alpha}^A \bar \psi_{\dot\alpha A}\,,\nonumber \\
u^4: \qquad  \omega^{(4)}_2 &= V^4\,.
\end{align}
However, by implementing the MC equations, the Hilbert series reduces to
\begin{eqnarray}\label{N4C}
 P(\Omega^\bullet(\mathfrak{g}/\mathfrak{h},\mathfrak{h}';\mathbb C),t,t^2) =1\,,
\end{eqnarray}
signalling the absence of nontrivial cohomology classes, as it can be confirmed by the following relations that \eqref{newOF} indeed satisfy:
\begin{eqnarray}
 \label{N4D}
 &&\diff \omega^{(3)} = \omega^{(4)}_1\,, ~~~~ \diff \omega^{(4)}_1 = 0\,, ~~~~~~~~\nonumber \\
 &&\diff \omega^{(4)}_2 = \omega^{(5)}\,, ~~~~ \diff \omega^{(5)} = 0\,.
\end{eqnarray}

In $D=4$ the supermultiplet for $N=4$ and $N=3$ super Yang-Mills coincide; therefore their analyses are expected to be related. On the other hand, we do not expect such relation in the supergravity cases, thus we briefly comment on this for the sake of completeness.

\paragraph{Comments on $N=3$ supersymmetry.} In this case, the spinors $\psi^A_\alpha$ and $\bar\psi_{A \dot \alpha}$ transform in the representation $3$ and $\bar 3$ of $SU(3)$. By computing the Molien-Weyl formula for $SO(4)$ invariant quantities, one gets the following series (denoting by $z_1$ and $z_2$ the parameters associated with $SU(3)$ and considering $\mathfrak{g}/\mathfrak{h}=(\mathfrak{su}(3)\ltimes \mathfrak{siso}(3|4))/(\mathfrak{so}(4)\oplus \mathfrak{su}(3))$ and $\mathfrak{so}(4)$):
\begin{align}\label{N3AA}
&P(\Omega^\bullet(\mathfrak{g}/\mathfrak{h},\mathfrak{h}';\mathbb C),t,t^2,z_1, z_2)=
1+\frac{\left(z_2 z_1^2+z_1^2+z_2^2 z_1+z_1+z_2^2+z_2\right)}{z_1 z_2} t^2 \nonumber \\
&+\frac{\left(z_2^2 z_1^4+z_1^4+z_2^2 z_1^3+z_2 z_1^3+z_2^4 z_1^2+z_2^3 z_1^2+z_2 z_1^2+z_1^2+z_2^3 z_1+z_2^2 z_1+z_2^4+z_2^2\right)}{z_1^2 z_2^2} t^4 +O\left(t^5\right) 
\end{align}
at order $t^4$.
The second term is easily understood, since there are two independent sets of cohomology representatives that we can build, namely
\begin{eqnarray}
\label{N3AB}
\omega^{(2) [AB]} = \epsilon^{\alpha\beta} \psi^A_\alpha \psi^B_\beta\,, ~~~
\bar\omega^{(2) }_{[AB]} = \epsilon^{\dot\alpha\dot\beta} \bar\psi_{A \dot\alpha}\bar \psi_{B \dot\beta}\,, ~~~
\end{eqnarray}
which transform in the representations $3$ and $\bar 3$. This is due to the properties of the $SU(3)$ representations, $3 \wedge 3 \sim \bar 3$. Indeed, the expression in \eqref{N3AA} proportional to 
$t^2$ is the sum of the characters of representations $3$ and $\bar 3$, which read  
\begin{eqnarray}
\label{N3AC}
\chi_\psi(z_1, z_2) = z_2+\frac{z_1}{z_2}+\frac{1}{z_1}\,, ~~~~
\chi_{\bar\psi}(z_1, z_2) = z_1+\frac{z_2}{z_1} +\frac{1}{z_2}\,.
\end{eqnarray}
Computing the third term in the expansion of \eqref{N3AA}, we get 
the following character: 
\begin{eqnarray}
\label{N3AD}
\frac{z_1^2}{z_2^2}+z_1^2+\frac{z_1}{z_2}+z_1+\frac{z_2^2}{z_1^2}+z_2^2+z_2+\frac{1}{z_1^2}+\frac{1}{z_2}+\frac{1}{z_2^2}+\frac{z_2}{z_1}+\frac{1}{z_1} \,,
\end{eqnarray}
which can be rewritten as 
\begin{eqnarray}
\label{N3AE}
\chi_\psi(z_1, z_2)^2 + \chi_{\bar\psi}(z_1, z_2)^2 - \chi_{\bar\psi}(z_1, z_2) - \chi_\psi(z_1, z_2) \,,
\end{eqnarray}
corresponding to the invariants 
\begin{eqnarray}
\label{N3AF}
\omega^{(2) [AB]}\wedge \omega^{(2) [CD]}\,, ~~~~~  \bar\omega^{(2) }_{[AB]} \wedge \bar\omega^{(2) }_{[CD]}\,, ~~~~~
\bar\omega^{(2) }_{[AB]} \wedge \omega^{(2) [CD]}\,, ~~~~~    {\omega^{(3)A}}_{B} = \psi^A_\alpha V^{\alpha \dot{\alpha}} \bar\psi_{B \dot{\alpha}} \,.
\end{eqnarray}
The last two invariants are related by acting with $\nabla$ on ${\omega^{(3)A}}_{B}$. In addition, from the first two invariants 
one has to subtract the reducible parts denoted by $- \chi_{\bar\psi}(z_1, z_2) - \chi_\psi(z_1, z_2)$ in the character. Therefore, 
the Molien-Weyl formula provides the complete spectrum of $R$-symmetry covariant expressions from which one can easily read those corresponding to cohomology representatives.
Let us mention, here, that the physical relevance of such covariant expressions should be understood in the framework of \emph{gauge hierarchies} \cite{menoMario}, whose analysis under the Molien-Weyl perspective will be subject of future investigations. 

\subsection{Curved cases}

We now focus on the study of cohomologies related to supergravities with negative cosmological constant, with various degrees of supersymmetry. 

\subsubsection{$N=1$ supersymmetry}

Let us start with the simplest curved supercoset $OSp(1|4)/SO(4)$\footnote{This is clearly not the only possible case, as one could, for instance, study $SU(2,2|1)/{\rm span}(\mathfrak{iso}(4)_K \oplus \mathfrak{u}(1) \oplus \mathcal{D} \oplus S)$, where $S$ are eight superconformal generators, $\mathcal{D}$ is the dilatation generator and $\mathfrak{u}(1)$ is the generator of the remaining $U(1)$ factor. The Lie algebra $\mathfrak{iso}(3,1)_K$ includes the Lorentz transformations $SO(3,1)$ and the 4 generators of the special conformal transformations.}, whose supervielbein satisfies the following MC equations:
{\begin{align}\label{MC1curv}
    &\nabla V^{\alpha\dot\alpha}=\bar\psi^{\dot\alpha}\psi^\alpha \,, \qquad \nabla\psi_\alpha=V_{\alpha\dot\alpha}\bar\psi^{\dot\alpha} \,, \qquad \nabla\bar\psi^{\dot\alpha}=V^{\alpha\dot\alpha}\psi_{\alpha} \,.
\end{align}}
As the Lorentz group is unchanged with respect to the flat case, we can still work with two component spinors. However, the covariant differential $\nabla$ now contains the spin connection and it is only nilpotent when acting on polynomials invariant under the whole subgroup $H$, which we are quotienting by. The characters and plethystic polynomials associated with the supervielbein are as in \eqref{N1B} and \eqref{pleA}. The only different ingredient, compared to the flat case, is the choice of scaling weight relation $u(t)=t$. Moreover, as discussed in Sec. \ref{workplan}, {for consistency one has to} attribute a weight to the differential $[\nabla]=t$. The Hilbert-Poincaré polynomial then reads (we are considering $\mathfrak{g}/\mathfrak{h}=\mathfrak{osp}(1|4)/\mathfrak{so}(4)$ and $\mathfrak{h}'=\mathfrak{h}=\mathfrak{so}(4)$)
\begin{eqnarray}
 \label{CC1A}
 P(\Omega^{\bullet}(\mathfrak{g}/\mathfrak{h}, \mathfrak{h}' ; \mathbb C ),t,t)=1 - t^3 + 3 t^4 - t^5\,, 
\end{eqnarray}
which corresponds to the following invariants:
\begin{eqnarray}\nonumber\label{CC1B}
    - t^3: \qquad \omega^{(3)} &=& V^{\alpha\dot\alpha} \psi_{\alpha} \bar \psi_{\dot\alpha}\,, \\ \nonumber
    3 t^4 : \qquad \omega^{(4)}_1 &=& V^{\alpha\beta}_2 \psi_{\alpha} \psi_{\beta}\,, ~~~~~~
\omega^{(4)}_2 = V_2^{\dot\alpha\dot\beta} \bar\psi_{\dot\alpha} \bar\psi_{\dot\beta}\,, \qquad \omega^{(4)}_3 = V_4\,,
\\ \nonumber
    - t^5: \qquad \omega^{(5)} &=& V_3^{\alpha\dot\alpha} \psi_{\alpha} \bar\psi_{\dot\alpha}\,.
\end{eqnarray}
We will now show how these polynomials are related by the differential: 
\begin{align}
    \nabla\omega^{(3)}=2\left(\omega^{(4)}_1-\omega^{(4)}_2\right) \,,\qquad \omega^{(5)}=\frac{1}{3!}\nabla\omega^{(4)}_1 \,,\qquad \nabla\left(\frac{1}{3!}\omega^{(4)}_1+\frac{1}{3!} \omega^{(4)}_2-2\omega^{(4)}_3\right)=0 \,.
\end{align}
The first relation shows that $\omega^{(3)}$ is not closed, the second one shows that $\omega^{(5)}$ is exact, whereas the last one shows that $\omega^{(4)}=\frac{1}{3!}\omega^{(4)}_1+\frac{1}{3!} \omega^{(4)}_2-2\omega^{(4)}_3$ is closed, but not exact. The latter is therefore the only nonexact cocycle among the invariants written above. Then the Hilbert-Poincaré polynomial for the cohomology reads $1+t^4$.\footnote{{Let us notice that the obtained result is consistent with theorems by Fuks and Greub-Halperin-Vanstone \cite{greub,Fuks}, which allows 
us to write the Hilbert series for coset spaces, in the case of equal-rank algebras.}} \\
To obtain the FDA structure, we should introduce a new $3$-form accounting for the cocycle $\diff A^{(3)}=\omega^{(4)}$ and consider a new FDA polynomial
\begin{eqnarray}\label{CCnew}
 P_{\text{FDA}}(\Omega^{\bullet}(\mathfrak{g}/\mathfrak{h}, \mathfrak{h}' ; \mathbb C ),t,t)=(1+t^4)(1-t^4) = 1-t^8\,.
\end{eqnarray}
The new resulting cohomology class is $\omega^{(7)}=A^{(3)} \wedge \omega^{(4)}$. We can then add a new (commuting) 6-form $A^{(6)}$ such that its MC equation reads
\begin{eqnarray}
    \diff A^{(6)} = \omega^{(7)} \,.
\end{eqnarray}
Thus, the Hilbert series is completely trivialized.

\subsubsection{$N=2$ supersymmmetry}

We again increase the amount of supersymmetry and consider the supercoset space $ OSp(2|4)/(SO(4) \times SO(2))$. The associated supervielbein satisfies the following MC equations:
\begin{align}\label{N2curvA}
    &\nabla V^{\alpha\dot\alpha}=\bar\psi^{\dot\alpha}_A\psi^{\alpha A} \,, \qquad \nabla\psi_\alpha^A=V_{\alpha\dot\alpha}\bar\psi^{\dot\alpha A} \,, \qquad \nabla\bar\psi^{\dot\alpha}_A=V^{\alpha\dot\alpha}\psi_{\alpha A} \,,
\end{align}
where the covariant derivative $\nabla$ may contain both spin connection and $SO(2)$ gauge connection, depending on the field it is acting on. The associated Hilbert-Poincaré series reads (we are taking $\mathfrak{g}/\mathfrak{h}=\mathfrak{osp}(2|4)/(\mathfrak{so}(4)\oplus \mathfrak{so}(2))$ and $\mathfrak{h}'=\mathfrak{h}={so}(4)\oplus \mathfrak{so}(2)$)
\begin{align}
    P(\Omega^\bullet(\mathfrak{g}/\mathfrak{h},\mathfrak{h}';\mathbb C),t,u) =\frac{1-2 t^2 \left(u^2-u+1\right) u+u^4}{\left(1-t^2\right)^2}\,,
\end{align}
which corresponds to the following generator invariants: 
\begin{align}
   \frac{1}{(1-t^2)^2}:\qquad &\omega^{(2)}_1=\epsilon_{AB}\epsilon^{\alpha\beta}\psi_\alpha^A\psi_\beta^B\,,\quad \omega^{(2)}_2=\epsilon^{AB}\epsilon^{\dot\alpha\dot\beta}\bar\psi_{\dot\alpha A}\bar\psi_{\dot\beta B} \,,\nonumber \\
   -2t^2u:\qquad &\omega^{(3)}_1=V^{\alpha\dot\alpha}\psi_{\alpha A}\bar\psi_{\dot\alpha B}\delta^{AB}\,, \quad \omega^{(3)}_2=V^{\alpha\dot\alpha}\psi_{\alpha A}\bar\psi_{\dot\alpha B}\epsilon^{AB} \,, \nonumber \\
   +2t^2u^2+u^4:\qquad & \omega^{(4)}_1=V_2^{\alpha\beta}\psi_{\alpha A}\psi_{\beta B}\delta^{AB}\,,\quad \omega^{(4)}_2=V_2^{\dot\alpha\dot\beta}\bar\psi_{\dot\alpha A}\bar\psi_{\dot\beta B}\delta^{AB}\,, \quad \omega^{(4)}_3=V_4 \,, \nonumber \\
   -2t^2u^3:\qquad &\omega^{(5)}_1=V_3^{\alpha\dot\alpha}\psi_{\alpha A} \bar\psi_{\dot\alpha B}\delta^{AB}\,,\quad \omega^{(5)}_2=V_3^{\alpha\dot\alpha}\psi_{\alpha A} \bar\psi_{\dot\alpha B}\epsilon^{AB} \,.
\end{align}
When the MC equations hold, we are left with
\begin{align}
    P(\Omega^\bullet(\mathfrak{g}/\mathfrak{h},\mathfrak{h}';\mathbb C),t,t)= \frac{1-2 t^3+3 t^4 -2 t^5}{\left(1-t^2\right)^2}= 1+2t^2-2t^3+6t^4-6t^5+O\left(t^6\right)
\end{align}
and the above generators are in general no longer independent. For instance, $\omega^{(5)}_2={\frac{1}{12}\omega^{(3)}_1\wedge\omega^{(2)}_2}$. We find the following relations:
\begin{align}\nonumber\label{relations}
    &\nabla\left(\omega^{(2)}_1-\omega^{(2)}_2\right)=0,\quad \frac14\nabla\left(\omega^{(2)}_1+\omega^{(2)}_2\right)=\omega^{(3)}_2\,,\quad \nabla\omega^{(3)}_1=\frac12\omega^{(2)}_1
\wedge\omega^{(2)}_2+2\left(\omega^{(4)}_1-\omega^{(4)}_2\right)\,,\\
&\nabla\left(\omega^{(4)}_3-\frac{1}{12}\left(\omega^{(4)}_1+\omega^{(4)}_2\right)+\frac{1}{48}\omega^{(2)}_1\wedge\omega^{(2)}_2\right)=0\,,\quad \nabla\omega^{(4)}_3= \omega^{(5)}_{1}\,,\nonumber \\ &\omega_{2}^{(3)}\left(\omega^{(2)}_1+\omega^{(2)}_2\right)=\nabla\left(-24\omega^{(4)}_3+2\left(\omega^{(4)}_1+\omega^{(4)}_2\right)\right)\,.
\end{align}
These relations imply cancellations order by order in the polynomial. The resulting series properly selects the cohomologies and reduces to 
\begin{align}
    P(\Omega^\bullet(\mathfrak{g}/\mathfrak{h},\mathfrak{h}';\mathbb C),t,t) =\frac{1+t^4}{1-t^2} \,.
\end{align}
The two cohomologies described here can be read from \eqref{relations},
\begin{align}
    \tilde\omega^{(2)}:= \omega^{(2)}_1-\omega^{(2)}_2\,, \quad \tilde\omega^{(4)}:=\omega^{(4)}_3-\frac{1}{12}(\omega^{(4)}_1+\omega^{(4)}_2)+\frac{1}{48}\omega^{(2)}_1\wedge\omega^{(2)}_2 \,.
\end{align}
Notice that a priori one could also consider any power of $\omega^{(4)}$, but it is possible to show that they do not give rise to new cohomology classes, as powers of $\omega^{(4)}$ can be expressed as combinations of $\omega^{(4)}$ itself and powers of $\omega^{(2)}$. One can then add a $1$-form potential $A^{(1)}$, the well-known graviphoton, such that $\nabla A^{(1)}=\tilde\omega^{(2)}$ and a $3$-form $A^{(3)}$ satisfying $\nabla A^{(3)}=\tilde\omega^{(4)}$. The series then becomes
\begin{align}
     P_{\text{FDA}}(\Omega^\bullet(\mathfrak{g}/\mathfrak{h},\mathfrak{h}';\mathbb C),t,t) =\frac{1+t^4}{1-t^2}(1-t^2)(1-t^4)=1-t^8,
\end{align}
implying the presence of an additional $7$-form cocycle built in terms of these new potentials. The addition of a $6$-form potential completely trivializes the polynomial. 

\subsubsection{$N=4$ supersymmetry}

To end this section, let us now consider the curved supercoset $OSp(4|4)/(SO(4) \times SO(4))$. In this case, the $SU(4)$ $R$-symmetry group of the flat rigid case breaks down to $SO(4)$, due to the gravitino mass terms or, equivalently, to the presence of the cosmological constant. \\ 
The supervielbein satisfies the MC equations
\begin{align}\label{OS4A}
    \nabla V^{\alpha\dot\alpha}=\bar\psi^{\dot\alpha}_{A\dot A}\psi^{\alpha A \dot A}\,, \qquad \nabla\psi_\alpha^{A\dot A}=V_{\alpha\dot\alpha}\bar\psi^{\dot\alpha A\dot A} \,,  \qquad \nabla \bar\psi_{\dot \alpha}^{A\dot A}=V_{\alpha\dot\alpha}\psi^{\alpha A\dot A} \,, 
\end{align}
where $A=1,2$, $\dot A=1,2$ are the $R$-symmetry $\mathfrak{so}(4) \sim \mathfrak{su}(2) \times \mathfrak{su}(2)$ indices. The computation of the Hilbert-Poincar\'e series with $\mathfrak{g}/\mathfrak{h}=\mathfrak{osp}(4|4)/(\mathfrak{so}(4) \oplus \mathfrak{so}(4))$ and $\mathfrak{h}'=\mathfrak{h}=\mathfrak{so}(4) \oplus \mathfrak{so}(4)$ yields
\begin{align}\label{poincpolN4curvtu}
    P(\Omega^\bullet(\mathfrak{g}/\mathfrak{h},\mathfrak{h}';\mathbb C),t,u) &= 
\frac{1 -t^2 (u-1)^2 u-2 t^4 \left(u^3+u\right)-t^6 (u-1)^2 u + u^4 + 2 t^8 u^2}{\left(1-t^4\right)^4}\,,
\end{align}
corresponding to the following invariant generators:
\begin{align}\label{gen}
    \frac{1}{(1-t^4)^4}:\qquad &\omega^{(4)}_1=\epsilon_{\alpha\beta} \epsilon_{\gamma\delta}\epsilon_{AB}\epsilon_{CD}\epsilon_{\dot A\dot C}\epsilon_{\dot B\dot D}\psi^{\alpha A\dot A}\psi^{\beta B\dot B}\psi^{\gamma C\dot C}\psi^{\delta D\dot D}\,,\nonumber\\ &\omega^{(4)}_2=\epsilon_{\dot\alpha\dot\beta} \epsilon_{\dot\gamma\dot\delta}\epsilon_{AB}\epsilon_{CD}\epsilon_{\dot A\dot C}\epsilon_{\dot B\dot D}\bar\psi^{\dot \alpha A\dot A}\bar\psi^{\dot\beta B\dot B}\bar\psi^{\dot\gamma C\dot C}\bar\psi^{\dot\delta D\dot D}\,,\nonumber\\
    &\omega^{(4)}_3=\epsilon_{\alpha\beta} \epsilon_{\dot\gamma\dot\delta}\epsilon_{AB}\epsilon_{CD}\epsilon_{\dot A\dot C}\epsilon_{\dot B\dot D}\psi^{\alpha A\dot A}\psi^{\beta B\dot B}\bar\psi^{\dot\gamma C\dot C}\bar\psi^{\dot\delta D\dot D}\,,\nonumber \\ &\omega^{(4)}_4=\epsilon_{\alpha\beta} \epsilon_{\dot\gamma\dot\delta}\epsilon_{AC}\epsilon_{BD}\epsilon_{\dot A\dot B}\epsilon_{\dot C\dot D}\psi^{ \alpha A\dot A}\psi^{\beta B\dot B}\bar\psi^{\dot\gamma C\dot C}\bar\psi^{\dot\delta D\dot D}\,,\nonumber\\
    -t^2u: \qquad & \omega^{(3)} = V^{\alpha \dot \alpha} \psi_\alpha^{A \dot A} \bar \psi_{\dot \alpha A \dot A} \,, \nonumber \\
    2t^2u^2: \qquad & \omega^{(4)}_5 = V_2^{\alpha \beta} \psi_{\alpha A \dot A} \psi_{\beta}^{A \dot A} \,, \quad \omega^{(4)}_6 = V_2^{\dot \alpha \dot \beta} \bar \psi_{\dot \alpha A \dot A} \bar \psi_{\dot \beta}^{A \dot A} \,, \nonumber \\
    -t^2u^3: \qquad & \omega^{(5)}_1 = V_3^{\alpha \dot \alpha} \psi_{\alpha A \dot A} \bar \psi_{\dot \alpha}^{A \dot A} \,, \nonumber \\
    -2t^4u: \qquad & \omega^{(5)}_2=V^{\alpha \dot \alpha} \psi^{A \dot A}_\alpha \bar \psi_{\dot \alpha}^{B \dot B} \psi_\beta^{C \dot C} \psi_\gamma^{D \dot D} \epsilon^{\beta \gamma} \epsilon_{AB} \epsilon_{CD} \epsilon_{\dot A \dot C} \epsilon_{\dot B \dot D} \ ,   \nonumber \\
    &\omega^{(5)}_3=V^{\alpha \dot \alpha} \psi^{A \dot A}_\alpha \bar \psi_{\dot \alpha}^{B \dot B} \bar\psi_{\dot\beta}^{C \dot C} \bar\psi_{\dot\gamma}^{D \dot D} \epsilon^{\dot\beta \dot\gamma} \epsilon_{AB} \epsilon_{CD} \epsilon_{\dot A \dot C} \epsilon_{\dot B \dot D} \ ,  \nonumber \\
    -2t^4u^3: \qquad &\omega^{(7)}_1=V_3^{\alpha\dot\alpha}\psi_{\alpha}^{A\dot A}\bar\psi_{\dot\alpha}^{B\dot B}\psi_{\beta}^{C\dot C}\psi_\gamma^{D\dot D}\epsilon^{\beta \gamma} \epsilon_{AB} \epsilon_{CD} \epsilon_{\dot A \dot C} \epsilon_{\dot B \dot D} \ ,   \nonumber \\ 
    &\omega^{(7)}_2=V_3^{\alpha\dot\alpha}\psi_{\alpha}^{A\dot A}\bar\psi_{\dot\alpha}^{B\dot B}\bar\psi_{\dot\beta}^{C\dot C}\bar\psi_{\dot\gamma}^{D\dot D}\epsilon^{\dot\beta \dot\gamma} \epsilon_{AB} \epsilon_{CD} \epsilon_{\dot A \dot C} \epsilon_{\dot B \dot D} \ ,   \nonumber \\ 
    -t^6u: \qquad &\omega^{(7)}_3=V^{\alpha\dot\alpha}\psi_{\alpha}^{A\dot A}\bar\psi_{\dot\alpha}^{B\dot B}\psi^{C\dot C}_\beta\psi^{D\dot D}_\gamma\bar\psi^{E\dot E}_{\dot\beta}\bar\psi^{F\dot F}_{\dot\gamma}\epsilon^{\beta\gamma}\epsilon^{\dot\beta\dot\gamma}\epsilon_{AC}\epsilon_{BE}\epsilon_{DF}\epsilon_{\dot A\dot C}\epsilon_{\dot B\dot E}\epsilon_{\dot D\dot F} \,, \nonumber \\
    2t^6u^2: \qquad & \omega^{(8)}_1=V_2^{\alpha\beta}\psi_\alpha^{A\dot A}\psi_\beta^{B\dot B}\psi_\gamma^{C\dot C}\psi_\delta^{D\dot D}\bar\psi_{\dot\beta}^{E\dot E}\bar\psi_{\dot\gamma}^{F\dot F}\epsilon^{\gamma\delta}\epsilon^{\dot\beta\dot\gamma}\epsilon_{AC}\epsilon_{BE}\epsilon_{DF}\epsilon_{\dot A\dot C}\epsilon_{\dot B\dot E}\epsilon_{\dot D\dot F}\,, \nonumber \\
    &\omega^{(8)}_2={V_2^{\dot\alpha\dot\beta}\bar\psi^{A\dot{A}}_{\dot{\alpha}}\bar\psi^{B\dot{B}}_{\dot{\beta}}\psi_\alpha^{C\dot C}\psi_\beta^{D\dot D}\bar\psi_{\dot\gamma}^{E\dot E}\bar\psi_{\dot\delta}^{F\dot F}\epsilon^{\alpha\beta}\epsilon^{\dot\gamma\dot\delta}\epsilon_{AC}\epsilon_{BE}\epsilon_{DF}\epsilon_{\dot A\dot C}\epsilon_{\dot B\dot E}\epsilon_{\dot D\dot F}} \,, \nonumber \\
    -t^6u^3: \qquad & \omega^{(9)} = {V_3^{\alpha \dot \alpha} \psi_{\alpha}^{A\dot A}\bar\psi_{\dot\alpha}^{B\dot B}\psi^{C\dot C}_\beta\psi^{D\dot D}_\gamma\bar\psi^{E\dot E}_{\dot\beta}\bar\psi^{F\dot F}_{\dot\gamma}\epsilon^{\beta\gamma}\epsilon^{\dot\beta\dot\gamma}\epsilon_{AC}\epsilon_{BE}\epsilon_{DF}\epsilon_{\dot A\dot C}\epsilon_{\dot B\dot E}\epsilon_{\dot D\dot F}} \,, \nonumber \\
    2t^8u^2: \qquad & \omega^{(10)}_1 = V_2^{\alpha \beta} \psi^{A\dot{A}}_\alpha \psi^{B\dot{B}}_\beta \bar\psi^{C\dot{C}}_{\dot\gamma} \bar\psi^{D\dot{D}}_{\dot\delta} \psi^{E\dot{E}}_{{\gamma}}  \psi^{F\dot{F}}_{{\delta}} \bar \psi^{L\dot{L}}_{\dot{\alpha}} \bar \psi^{M\dot{M}}_{\dot{\beta}}\times \nonumber \\
    &\qquad \qquad \qquad\qquad \times\epsilon^{\dot\gamma \dot\delta} \epsilon^{{\gamma}{\delta}} \epsilon^{\dot\alpha\dot\beta} \epsilon_{AB}\epsilon_{\dot B\dot C} \epsilon_{CD} \epsilon_{\dot D\dot E} \epsilon_{EF} \epsilon_{\dot{F}\dot{L}} \epsilon_{LM}
    \epsilon_{\dot{M}\dot{A}}  \,, \nonumber \\
    & \omega^{(10)}_2 = V_2^{\dot\alpha \dot\beta} \bar\psi^{A\dot{A}}_{\dot{\alpha}} \bar\psi^{B\dot{B}}_{\dot{\beta}} \psi^{C\dot{C}}_\gamma \psi^{D\dot{D}}_\delta \bar \psi^{E\dot{E}}_{\dot{\gamma}} \bar \psi^{F\dot{F}}_{\dot{\delta}}  \psi^{L\dot{L}}_{\alpha}  \psi^{M\dot{M}}_{\beta}\times\nonumber\\
    &\qquad \qquad \qquad\qquad \times\epsilon^{\gamma\delta} \epsilon^{\dot{\gamma}\dot{\delta}} \epsilon^{\alpha\beta} \epsilon_{AB}\epsilon_{\dot B\dot C} \epsilon_{CD} \epsilon_{\dot D\dot E} \epsilon_{EF} \epsilon_{\dot{F}\dot{L}} \epsilon_{LM}
    \epsilon_{\dot{M}\dot{A}}  \,, \nonumber \\
    u^4: \qquad & \omega^{(4)}_7 = V_4 \,.
\end{align}
When the MC equations hold, the generators in \eqref{gen} are no longer independent. As explained in the $N=2$ case, this implies order by order cancellations among the invariant polynomials such that the resulting series counting cohomology classes reads 
\begin{align}\label{OS4B}
P(\Omega^\bullet(\mathfrak{g}/\mathfrak{h},\mathfrak{h}';\mathbb C),t,t) &= {\frac{1+t^4}{(1-t^4)^2}} \,.
\end{align}
In particular, from the numerator, we see that there are two 4-form cocycles, $\omega^{(4)}_1$ and $\omega^{(4)}_2$, which can then be trivialized to build the FDA by introducing $A^{(3)}_1$ and $A^{(3)}_2$ ($\diff A^{(3)}_1=\omega^{(4)}_1$, $\diff A^{(3)}_1=\omega^{(4)}_2$). This procedure modifies \eqref{OS4B} as 
\begin{eqnarray}
\label{OS4E}
P_{\text{FDA}}(\Omega^\bullet(\mathfrak{g}/\mathfrak{h},\mathfrak{h}';\mathbb C),t,t) = \frac{1+t^4}{(1-t^4)^2} (1-t^4)^2 = 1+t^4 \,,
\end{eqnarray}
which shows that the introduction of the 3-forms $A^{(3)}_1$ and $A^{(3)}_2$ does not trivialize the FDA yet. In particular, we have a 4-form cocycle $\omega^{(4)}_3$, whose trivialization requires a new 3-form $A^{(3)}_3$ such that $\omega^{(4)}=\diff A^{(3)}$. 
Including the 3-form $A^{(3)}_3$, the FDA series becomes
\begin{eqnarray}
\label{OS4Enew}
P'_{\text{FDA}}(\Omega^\bullet(\mathfrak{g}/\mathfrak{h},\mathfrak{h}';\mathbb C),t,t) = ({1+t^4})({1-t^4}) = 1-t^8 \,.
\end{eqnarray}
Again, the FDA is not trivialized yet, as we are left with a 7-form cocycle $\omega^{(7)}$, which can be trivialised by a new even 6-form generator $B^{(6)}$, with weight $t^8$, such that
\begin{equation}
    \diff B^{(6)} = \omega^{(7)} \,.
\end{equation}
This procedure completely trivializes the Hilbert-Poincar\'e polynomial
\begin{equation}
   P''_{\text{FDA}}(\Omega^\bullet(\mathfrak{g}/\mathfrak{h},\mathfrak{h}';\mathbb C),t,t) = \frac{1-t^8}{1-t^8} = 1 \,.
\end{equation}

\section{$D=6$ spacetime dimensions}\label{sec4}

In this section, we move on to the study of cocycles and FDAs for various vacuum supergravity theories in $D=6$, with different amount of supersymmetry.

\subsection{Flat rigid superspaces}

Let us focus, here, on different flat rigid superspace cases, in which the Molien-Weyl formula directly selects all the cohomologies out of the invariant polynomials computed by the Hilbert-Poincaré series.

\subsubsection{$N=(4,0)$ supersymmetry}\label{secD6N40}

We start by computing the relevant cohomology classes for the flat $D=6$, $N=(4,0)$ case. In the literature, different nomenclatures are used for the various six-dimensional supergravity theories. In particular, by $N=(4,0)$, which is also known in ``American'' notation as $N=(2,0)$ theory, we denote the chiral model with $R$-symmetry group $USp(4)\sim SO(5)$, whose real dimension is 10.
The computation of the Molien-Weyl integration formula simplifies if we choose the following convenient representation of $SO(6)\times USp(4)$ onto to coset space {$(USp(4)\ltimes sISO((4,0)|6))/(SO(6)\times USp(4))$}:
\begin{eqnarray}
\label{newA}
V_{[\alpha \beta]}\,,~~~~~ \psi^A_\alpha\,, \end{eqnarray}
where the index $A=1,\ldots,4$ runs over the $USp(4)$ representation (with symplectic structure  $\mathbb C_{AB}$) and the indices $\alpha, \beta,\ldots=1, \ldots,4$ run over the $SU(4)$ fundamental representation. \\
The MC equations are
\begin{eqnarray}
 \label{newB}
 \nabla V_{[\alpha \beta]} = \psi^A_\alpha \psi^B_\beta \mathbb C_{AB}\,, ~~~~~
 \nabla \psi^A_\alpha =0\,, 
\end{eqnarray}
with $\nabla$ nilpotent. We now introduce the various ingredients, needed for the computation of the Hilbert-Poincaré polynomial: to this end, let us consider the characters of the $SU(4)$ antisymmetric representation of $V_{[\alpha\beta]}$ and 
of the spinors $\psi_\alpha^A$, which read, respectively,
\begin{eqnarray}
 \label{newG}
 \chi_V(z_1,z_2,z_3) &=& z_2 + \frac{z_1 z_2}{z_3} + \frac{z_3}{z_1} 
 +\frac{z_1}{z_3} +   \frac{z_3 }{z_1 z_2} + \frac{1}{z_2} \,, \nonumber \\ 
 \chi_\psi(z_1,z_2,z_3, z,w) &=& 
 \left(z_1 + \frac{z_2}{z_1} + \frac{z_3}{z_2} + \frac{1}{z_3}\right) 
 \left(w + \frac1w + z + \frac1z \right) \,.
\end{eqnarray}
Here $z$ and $w$ parametrize the two torus $U(1)\times U(1)$ for the two Cartan generators of $USp(4)$, while $z_1,z_2,z_3$ are the parameters of the three torus $U(1)\times U(1)\times U(1)$ of $SU(4)$. The associated plethystic polynomials read
\begin{eqnarray}
 \label{newF}
 PE[\chi_Vu] &=& \left(1-u z_2\right) \left(1-\frac{u z_1 z_3}{z_2}\right) \left(1-\frac{u z_3}{z_1}\right) \left(1-\frac{u z_2}{z_1 z_3}\right) \left(1-\frac{u z_1}{z_3}\right) \left(1-\frac{u}{z_2}\right) \,, \nonumber \\
 PE[\chi_\psi t] &=& \frac{1}{\prod_{i=1}^4 \Delta_i} \,, \quad \text{with} \nonumber \\
 \Delta_1 &=&\left(1-t w z_1\right) \left(1-\frac{t z_1}{w}\right) \left(1-t z z_1\right) \left(1-\frac{t z_1}{z}\right) \,, \nonumber \\
 \Delta_2 &=& \left(1-\frac{t z_2}{w z_1}\right) \left(1-\frac{t w z_2}{z_1}\right) \left(1-\frac{t z_2}{z z_1}\right) \left(1-\frac{t z z_2}{z_1}\right) \,, \nonumber \\
 \Delta_3 &=& \left(1-\frac{t w z_3}{z_2}\right) \left(1-\frac{t z_3}{w z_2}\right) \left(1-\frac{t z z_3}{z_2}\right) \left(1-\frac{t z_3}{z z_2}\right) \,, \nonumber \\
 \Delta_4 &=&  \left(1-\frac{t w}{z_3}\right) \left(1-\frac{t}{w z_3}\right) \left(1-\frac{t z}{z_3}\right) \left(1-\frac{t}{z z_3}\right) \,.
\end{eqnarray}
At last, let us consider the integration measure of $USp(4)$
\begin{eqnarray}
 \label{newD}
 \diff \mu_{USp(4)} = (1 - w^2)(1-z^2) (1 - w z) \left(1 - \frac{w}{z}\right) \frac{dz dw}{(2 \pi i)^2 z w}
\end{eqnarray}
and that of $SU(4)$, given in \eqref{newE1}. \\
The Hilbert-Poincaré series for $\mathfrak{g}/\mathfrak{h}=(\mathfrak{usp}(4)\ltimes \mathfrak{siso}((4,0)|6))/(\mathfrak{so}(6)\oplus \mathfrak{usp}(4))$ and $\mathfrak{h}'=\mathfrak{h}=\mathfrak{so}(6)\oplus \mathfrak{usp}(4)$ then reads
\begin{eqnarray}
 \label{newH}
 P(\Omega^\bullet(\mathfrak{g}/\mathfrak{h},\mathfrak{h}';\mathbb C),t,u)  = \frac{1 - t^2 (u + u^5) + u^6}{1 - t^4} \,.
\end{eqnarray}
The different elements correspond to the following invariant expressions:
\begin{align}\label{newL}
\frac{1}{1-t^2} : \qquad & \omega^{(4)} = \psi^A_{\alpha} \psi^B_{\beta} \psi^C_{\gamma} \psi^D_{\delta} \epsilon^{\alpha\beta\gamma\delta} \mathbb C_{AB} \mathbb C_{CD} \,, \nonumber \\
 -t^2 u: \qquad & \omega^{(3)}  = V_{\alpha \beta} 
 \psi^C_{\gamma} \psi^D_{\delta} \mathbb C_{CD}\epsilon^{\alpha\beta\gamma\delta} \,, \nonumber \\
 -t^2 u^5: \qquad & \omega^{(7)} = {V_5^{[\alpha\beta]}\psi_\alpha^C\psi_\beta^D\mathbb C_{CD}} \,, \nonumber \\
u^6: \qquad & \omega^{(6)} =V_6 \,,
\end{align}
{where we defined}
\begin{align}\nonumber
    V_{5}^{[\alpha \beta]}&=\frac{1}{5!}
  V_{\alpha_1 \beta_1} V_{\alpha_2 \beta_2}V_{\alpha_3 \beta_3}V_{\alpha_4 \beta_4}V_{\alpha_5 \beta_5} 
  \epsilon^{\alpha_1 \alpha_2\beta_1\beta_2} \epsilon^{\alpha_3\alpha_4\alpha_5 {\alpha}} \epsilon^{\beta_3\beta_4{\beta_5}{\beta}}\,, \\ \nonumber
  V_6&=\frac{1}{6!}
   V_{\alpha_1 \beta_1} V_{\alpha_2 \beta_2}V_{\alpha_3 \beta_3}V_{\alpha_4 \beta_4}V_{\alpha_5 \beta_5}V_{\alpha_6\beta_6}\epsilon^{\alpha_1 \alpha_2\beta_1\beta_2} \epsilon^{\alpha_3\alpha_4\alpha_5 {\alpha_6}} \epsilon^{\beta_3\beta_4{\beta_5}{\beta_6}} \,.
\end{align}
Note that, due to antisymmetrizations, the combination $\mathbb C_{AB} \mathbb C_{CD}$ can be rewritten as $\epsilon_{ABCD} = 
\mathbb C_{AB} \mathbb C_{CD} - \mathbb C_{CB} \mathbb C_{AD}- \mathbb C_{AD} \mathbb C_{CB}$
and, therefore, there is a single nontrivial expression at level $1/(1-t^2)$. The 4-form $\omega^{(4)}$, is closed, not exact and invariant under Lorentz and $R$-symmetry transformations. However, by imposing the MC equations \eqref{newB}, thus setting $u=t^2$, we get 
\begin{equation}
    \label{newM}
    P(\Omega^\bullet(\mathfrak{g}/\mathfrak{h},\mathfrak{h}';\mathbb C),t,t^2) = 1 \,.
\end{equation}
indicating that there are no cohomology classes. This follows from the following relations
\begin{equation}\label{newN}
\diff \omega^{(3)}  = \omega^{(4)} \,, ~~~~ \diff \omega^{(6)}  = \omega^{(7)} \,,
\end{equation}
that show that the cohomology is indeed trivial.\\

However, to make contact with the associated supergravity theory (see, e.g., \cite{CDF,Andrianopoli:2022bzr}), we should take into account that the gravitational multiplet of this theory also includes five 2-forms $B^{AB}$ (in the 2-times antisymmetric traceless irrep of $USp(4)$) with selfdual field strengths, which also contribute to the FDA, through 
\begin{eqnarray}
\label{aggiuB}
dB^{AB}= {\psi^{[A}_\alpha\psi^{B]_0}_\beta V_{\gamma\delta}\epsilon^{\alpha\beta\gamma\delta}} \,,
\end{eqnarray}
where with $[AB]_0$ we denote tracelessness in the antisymmetric indices $[AB]$, that is $K_{[AB]_0}=K_{[AB]}-\frac 14 \mathbb{C}_{AB}\mathbb{C}^{CD}K_{CD}$. To recover this object with the Molien-Weyl formula, one has to compute the number of Lorentz-invariant polynomials only, i.e., without integrating on the $USp(4)$ parameters. Indeed, in this case ($\mathfrak{h}'=\mathfrak{so}(6)$), one gets
 \begin{eqnarray}
\label{aggiuA}
 P(\Omega^\bullet(\mathfrak{g}/\mathfrak{h},\mathfrak{h}';\mathbb C),t,t^2,z,w) = 1 - t^4 \left(1 +  w z + \frac{w}{z} + \frac{z}{w} + \frac{1}{w z}\right) \,,
\end{eqnarray}
which corresponds exactly to the invariant \eqref{aggiuB}, with scale $t^4$, in the traceless representation of $USp(4)$. 

\subsubsection{$N=(2,2)$ supersymmetry}

{Another $D=6$ flat rigid superspace case that we wish to study is the $N=(2,2)$ case, which is a non-chiral theory with $R$-symmetry group $SU(2)\times SU(2)$.} In this case, the associated coset space is $((SU(2) \times SU(2)) \ltimes sISO((2,2)|6))/(SO(6)\times SU(2)\times SU(2))$, which features two spinor fields with opposite chiralities. The corresponding MC equations read
\begin{eqnarray}\label{UAAA}
\nabla V_{[\alpha\beta]} = \psi^A_\alpha \psi^B_\beta \mathbb C_{AB} + \epsilon_{\alpha\beta\gamma\delta} \,
 \psi^{\bar A \gamma} \psi^{\bar B \delta} \mathbb C_{\bar A \bar B}\,, 
 ~~~~~ \nabla \psi^A_\alpha =0\,, ~~~~~\nabla \psi^{\bar A \alpha}=0\,, 
\end{eqnarray}
where $\nabla$ is a nilpotent operator and $\psi^A_\alpha$ are 8 spinors with left chirality (with $A,B,\ldots=1,2$), whereas $\psi^{\bar A \alpha}$ are the other 8 spinors with right chirality (with $\bar A , \bar{B},\ldots=1,2$). \\
The ingredients needed for the computation of the Molien-Weyl formula receive slight modifications with respect to the ones in Sec. \ref{secD6N40}. In particular, even if the character associated with $V_{[\alpha \beta]}$ remains unaltered, those of the spinors $\psi^A$ and $\psi^{\bar{A}}$ now read, respectively,
\begin{eqnarray}
 \label{newUAB}
 \chi_{\psi^A}(z_1,z_2,z_3,w) &=& 
 \left(z_1 + \frac{z_2}{z_1} + \frac{z_3}{z_2} + \frac{1}{z_3}\right)  \left(w + \frac1w\right) \,, \nonumber \\
 \chi_{{\psi^{\bar{A}}}}(z_1,z_2,z_3,z) &=& 
 \left(z_3 + \frac{z_2}{z_3} + \frac{z_1}{z_2} + \frac{1}{z_1}\right)
 \left(z + \frac1z\right) \,,
\end{eqnarray}
as the two spinors have opposite chirality-independent $R$-symmetry characters.
Moreover, the integration measure on the $R$-symmetry group now becomes the one given in \eqref{newUAA1}. 
Then, the computation of the Molien-Weyl formula for  $\mathfrak{g}/\mathfrak{h}=((\mathfrak{su}(2) \oplus \mathfrak{su}(2)) \ltimes \mathfrak{siso}((2,2)|6))/(\mathfrak{so}(6)\oplus \mathfrak{su}(2)\oplus \mathfrak{su}(2))$ and $\mathfrak{h}'=\mathfrak{h}=\mathfrak{so}(6)\oplus \mathfrak{su}(2)\oplus \mathfrak{su}(2)$ gives
\begin{eqnarray}
\label{newUAC}
P(\Omega^\bullet(\mathfrak{g}/\mathfrak{h},\mathfrak{h}';\mathbb C),t,u)= \frac{1 - 2 t^2 u + t^4 u^2 + t^4 u^4 - 2 t^2 u^5 + u^6}{1- t^4} \,.
\end{eqnarray}
We are now able to list all possibile invariants described by the Hilbert series, namely 
\begin{align}\label{UAC}
t^4 : \qquad & \omega^{(4)} = \mathbb C_{AB} \psi_\alpha^A \psi_\beta^B  \mathbb C_{\bar C\bar D} 
\psi^{\alpha \bar C}\psi^{\beta \bar D} \,, \nonumber \\   
-2 t^2 u : \qquad & \omega^{(3)}_1 = V_{[\alpha \beta]} \epsilon^{\alpha \beta \gamma\delta } \mathbb C_{AB}  \psi_\gamma^A \psi_\delta^B \,, \nonumber \\
& \omega^{(3)}_2=   V_{[\alpha \beta]} \mathbb C_{\bar A\bar B}  \psi^{\alpha \bar A} \psi^{\beta \bar B} \,, \nonumber \\
t^4 u^2 : \qquad & \omega^{(6)} =  \omega^{(3)}_{{1}} \wedge  \omega^{(3)}_{{2}} \,, \nonumber \\
t^4 u^4 : \qquad & \omega^{(8)}= V_4^{\alpha \beta \delta\gamma } 
\epsilon_{\delta\gamma  \sigma \rho} 
\left(\mathbb C_{AB}  \psi_\alpha^A \psi_\beta^B  \right) 
\left(\mathbb C_{\bar C\bar D}  \psi^{\rho \bar C} \psi^{\sigma \bar D}\right) \,, \nonumber \\
-2t^2 u^5 : \qquad &  \omega^{(7)}_1 = V_5^{[\alpha \beta]} \mathbb C_{AB}  \psi_\alpha^A \psi_\beta^B \,, \nonumber \\
& \omega^{(7)}_2 =   V_5^{[\alpha \beta]} 
 \epsilon_{\alpha \beta  \sigma \rho} \mathbb C_{\bar A\bar B}  \psi^{\sigma \bar A} \psi^{\rho \bar B} \,, \nonumber \\
u^6 : \qquad & \omega^{(6)} = V^6 \,,
\end{align}
where we defined $V_4^{\alpha\beta\gamma\delta} = \frac{1}{4!}V_{\alpha_1 \beta_1} V_{\alpha_2 \beta_2}V_{\alpha_3 \beta_3}V_{\alpha_4 \beta_4} 
  \epsilon^{\alpha_1 \alpha_2\beta_1\beta_2} \epsilon^{\alpha_3\alpha_4 {\alpha}\beta} \epsilon^{\beta_3\beta_4{\gamma}\delta}$.\\
The cocycles are then obtained by imposing the MC equations: by doing so, we get
\begin{eqnarray}\label{newUAD}
P(\Omega^\bullet(\mathfrak{g}/\mathfrak{h},\mathfrak{h}';\mathbb C),t,t^2)= 1 - t^4\,, 
\end{eqnarray}
which implies that the only cohomology class is 
\begin{eqnarray}
\label{UAE}
\tilde{\omega}^{(3)} = \omega^{(3)}_1 - {4}\omega^{(3)}_2 = 
V_{[\alpha \beta]}\left( 
\epsilon^{\alpha \beta \gamma\delta } \mathbb C_{AB}  \psi_\delta^A \psi_\gamma^B - {4}
   \mathbb C_{\bar A\bar B}  \psi^{\alpha \bar A} \psi^{\beta \bar B} \right) \,,
\end{eqnarray}
as $\diff \omega^{(3)}_1 = {4}\diff \omega^{(3)}_2$, we have $\diff \tilde{\omega}^{(3)}=0$. The expression in \eqref{UAE} is the usual 3-form cocycle needed for {the Green-Schwarz Wess-Zumino term} by integrating it over $D=3$ supermanifold \cite{Green:1987sp}. \\
Notice that this model is crucially different form the $N=(4,0)$ we have previously considered, in which case, in fact, the 3-form $\omega^{(3)}$ in \eqref{newL} is not closed.
Here, in the $N=(2,2)$ model, the closure of $\tilde{\omega}^{(3)}$ is due to Fierz and Schouten identities. One can then add a (commuting) 2-form gauge potential $B^{(2)}$ such that
\begin{equation}
    \diff B^{(2)} = \tilde{\omega}^{(3)} \,,
\end{equation}
which trivializes the Hilbert polynomial,
\begin{equation}
    P_{\text{FDA}}(\Omega^\bullet(\mathfrak{g}/\mathfrak{h},\mathfrak{h}';\mathbb C),t,t^2)= \frac{1 - t^4}{1-t^4} = 1 \,. 
\end{equation}
Let us conclude our study by mentioning that, as far as the FDA of the associated $D=6$ supergravity theory is concerned, the supergravity multiplet should include, besides the supervielbein, the 2-form gauge potential trivializing $\tilde{\omega}^{(3)}$, the appropriate spin-$1/2$ fields and a scalar, also four gauge 1-forms such that
\begin{equation}
    \diff A^{A\bar B}= \psi^A_\alpha {\psi}^{\bar B \alpha} \,.
\end{equation}
The latter do not appear in the above discussion as they are not $R$-symmetry invariant. Really, in order to find these objects within our approach, one should consider $\mathfrak{h}'=\mathfrak{so}(6)$, which indeed allows the explicit computation of Lorentz-invariant expressions (and, in particular, cocycles) with free $R$-symmetry indices.

\subsubsection{$N=(2,0)$ supersymmetry}

Let us analyze here the $N=(2,0)$ case, which is a chiral theory with $SU(2)$ $R$-symmetry group, whose corresponding coset space is $(SU(2) \ltimes sISO((2,0)|6))/(SO(6)\times SU(2))$. \\
The character of the vector representation of the vielbein remains unaltered with respect to the one in Sec. \ref{secD6N40}, whereas the one of the gravitini becomes
\begin{eqnarray}\label{newO}
 \chi_\psi(z_1,z_2,z_3,w) &=& 
 \left(z_1 + \frac{z_2}{z_1} + \frac{z_3}{z_2} + \frac{1}{z_3}\right) 
 \left(w + \frac1w\right) \,.
\end{eqnarray}
In addition, the integration measure related to the $R$-symmetry group in the case at hand has to be changed with respect to the one in Sec. \ref{secD6N40} to the one given in \eqref{newE}. \\
The resulting Hilbert-Poincaré polynomial is
\begin{eqnarray}
 \label{newP}
 P(\Omega^\bullet(\mathfrak{g}/\mathfrak{h},\mathfrak{h}';\mathbb C),t,u) = 1 - t^2 (u + u^5) + u^6 \,,
\end{eqnarray}
where now $\mathfrak{g}/\mathfrak{h}=(\mathfrak{su}(2) \ltimes \mathfrak{siso}((2,0)|6))/(\mathfrak{so}(6)\oplus \mathfrak{su}(2))$ and $\mathfrak{h}'=\mathfrak{h}=\mathfrak{so}(6)\oplus \mathfrak{su}(2)$. Let us observe that the expression on the right-hand side of \eqref{newP} is just the numerator of the $N=(4,0)$ case, given in \eqref{newH}; the reason is that the expressions for the invariants in {\eqref{newL}} remain unaltered in the present case, except for the fact that $\mathbb C_{AB}$ is replaced by the symplectic matrix for $SU(2)$, which here, with a little abuse of notation, we denote in the same way as $\mathbb C_{AB}$. The denominator is absent, since there is no quartic invariant $\omega^{(4)}_1$ as in \eqref{newL}, which now vanishes, due to the following Schouten identity:
\begin{equation}
    \mathbb C_{AB} \mathbb C_{CD} - \mathbb C_{CB} \mathbb C_{AD}- \mathbb C_{AD} \mathbb C_{CB} = 0 \,,
\end{equation}
with $A,B,C,D=1,2$. Notice also that we could have easily guessed the expression of \eqref{newP} from that of \eqref{newUAC} by ``switching off'' the invariants built out of one of the two $SU(2)$ symplectic matrices $\mathbb C$. \\
By imposing the MC equations, we get 
\begin{eqnarray}\label{newQ}
 P(\Omega^\bullet(\mathfrak{g}/\mathfrak{h},\mathfrak{h}';\mathbb C),t,t^2)= 1 - t^4 \,,
\end{eqnarray}
indicating that there is a single cohomology class, corresponding to $\omega^{(3)}$ of Sec. \ref{secD6N40} (but where now $\mathbb C_{AB}$ denotes the symplectic matrix for $SU(2)$). In fact, this is the class needed to construct the Wess-Zumino term (D1-brane). In a purely supergravity context, this can be understood since the corresponding supergravity multiplet includes one antisymmetric 2-form $B^{(2)}$ with selfdual field strength, which is naturally interpreted as a trivialization of the 3-cycle $\omega^{(3)}$: $\diff B^{(2)}= \omega^{(3)}$.

\subsection{Curved case $N=2$, $AdS_3 \times S^3$}

In this section we explore the $D=6$, $N=2$ curved case, considering the supercoset $OSp(4|2)/SO(3)$. Its 
bosonic sector is described by the Cartesian product $AdS_3 \times S^3$ and it has 
eight supersymmetries. We use the following notation for the supervielbein: 
\begin{eqnarray}
\label{ad3_A}
V^{\alpha\beta}\,, ~~~~~ A_{ab}\,, ~~~~~ \psi^\alpha_{a \dot a} \,,
\end{eqnarray}
where $\alpha, \beta =1,2$ are $Sp(2)$ indices, $a,b=1,2$ are $SO(3) \subset SO(4)$ indices and $\dot a=1,2$ is an index of the gauged subgroup $SO(3) \subset SO(4)$. 
The associated MC equations are 
\begin{eqnarray}
\label{ad3_B}
{\nabla} V^{\alpha\beta} &=& \psi^\alpha_{a \dot a} \psi^\beta_{b \dot b} \epsilon^{ab} \epsilon^{\dot a\dot b} + (V\wedge V)^{\alpha\beta}\,, \nonumber \\
{\nabla} A_{a b} &=& \psi^\alpha_{a \dot a} \psi^\beta_{b \dot b} \epsilon_{\alpha \beta} \epsilon^{\dot a\dot b} + (A\wedge A)_{a b}\,, \nonumber \\ 
{\nabla} \psi^\alpha_{a \dot a} &=& V^{\alpha}_{~\beta} \psi^\beta_{a \dot a} + A_{a}^{~b} \psi^\beta_{b \dot a}\,.
\end{eqnarray}
In order to apply the Molien-Weyl formula, we need the characters
\begin{eqnarray}
\label{ads3_D}
\chi_{V}(z) &=& \left(  z^2+ 1 + \frac{1}{z^2}\right)  \,, \nonumber \\
\chi_{A}(w) &=& \left( w^2 + 1  + \frac{1}{w^2}\right)  \,, \nonumber \\
\chi_{\psi}(x,z,w) &=& \left( x w z + \frac{x w}{z}+
\frac{w z}{x} +  \frac{x z}{w} + \frac{ x }{ w z} +
\frac{z }{ w x}+ \frac{w }{ x z} + \frac{1}{x w z} \right) \,.
\end{eqnarray}
and the integral measure \eqref{newE}.
One can then prove that, setting $u(t)=t$, the following Hilbert series emerges (we are considering $\mathfrak{g}/\mathfrak{h}=\mathfrak{osp}(4|2)/\mathfrak{so}(3)$ and $\mathfrak{h}'=\mathfrak{h}=\mathfrak{so}(3)$): 
\begin{align}\label{ads3_E}\nonumber
P(\Omega^\bullet(\mathfrak{g}/\mathfrak{h},\mathfrak{h}';\mathbb C),t,t,z,w)&=\frac{(1-t^4)(1-z^2t)(1-t)(1-\frac{t}{z^2})(1-w^2t)(1-t)(1-\frac{t}{w^2})}{(1-t^2)^2(1-t^2w^2)(1-\frac{t^2}{w^2})(1-t^2z^2)(1-\frac{t^2}{z^2})}\\ \nonumber
&=1-t \left(\left(1 +\frac{1}{w^2}+ w^2\right) + \left(1 + \frac{1}{z^2} + z^2\right)\right)+\\ \nonumber
&+t^2\bigg(\left(1 +\frac{1}{w^2} + w^2\right) + \left(1 + \frac{1}{z^2} + 
   z^2\right) + \left(1 + \frac{1}{w^2} + w^2\right) \left(1 + \frac{1}{z^2} + z^2\right) \\
   &\qquad \ + 2 + w^2 + \frac{1}{w^2} +
 z^2 + \frac{1}{z^2}\bigg)+O(t^3) \,.
\end{align}
From the numerator in the first expression one can see that both $V$ and $A$ and any of their products are invariant; from the denominator one can see that the bilinears of the form $\psi^\alpha_{a \dot a} \psi^{\beta}_{b \dot b} \epsilon^{\dot a \dot b}$ and any of their products are invariant. Finally, the factor $(1-t^4)$ in the numerator keeps track of a Schouten identity for quadrilinears in $\psi$.\\
The computation of the cohomology classes is greatly simplified thanks to the above formula, which directly identifies $SO(3)$-invariant forms in the coset. The explicit construction of these objects can be performed analogously to what has been done for curved superspaces in previous sections.

\section{$D=10$ spacetime dimensions}\label{sec5}

\subsection{$N=1$ type I flat rigid superspace}

Let us now apply the Molien-Weyl formula to the $D=10$, $N=1$ superspace case, in order to compute the Lorentz-invariant cocycles. The coset space is $G/H=sISO(1|10)/SO(10)$, which can be dually described in terms of the supervielbein $(V^a,\psi^\alpha)$, satisfying the following MC equations:\footnote{Here $a,b,\ldots=0,1,\ldots,9$ are vector indices, while $\alpha=1, \ldots, 16$ is a spinorial index; we use the minimal irreducible representation, namely $\psi^\alpha$ is a Majorana-Weyl spinor in $D=10$. For simplicity, in the following, we will frequently omit writing the spinorial index.}
\begin{eqnarray}
\label{die0A}
{\diff} V^a = {\frac{i}{2}} {\bar{\psi}} \gamma^a \psi\,, ~~~~~~
{\diff} \psi =0\,,
\end{eqnarray}
where ${\diff}$ is nilpotent and the $\gamma^a$'s are {$16 \times 16$ gamma matrices} in ten dimensions. \\
The characters for the 10-dimensional vector representation and 16-dimensional spinorial representations of $SO(5)$, which has rank $5$ and corresponds to the Dynkin label $D_5$, read
\begin{eqnarray}
\chi_{V} (z_1,\ldots,z_5) &=& \frac{z_1}{z_2}+z_1+\frac{z_3}{z_2}+\frac{z_4 z_5}{z_3}+\frac{z_5}{z_4}+\frac{z_2}{z_3}+\frac{z_4}{z_5}+\frac{z_3}{z_4 z_5}+\frac{z_2}{z_1}+\frac{1}{z_1} \,, \label{dieB} \\
\chi_{\psi} (z_1,\ldots,z_5) &=&  \frac{z_4 z_1}{z_3}+\frac{z_5 z_1}{z_2}+\frac{z_1}{z_4}+\frac{z_3 z_1}{z_2 z_5}+\frac{z_4}{z_2}+z_4+\frac{z_2 z_5}{z_3}\nonumber \\
&+&\frac{z_5}{z_3}+\frac{z_3}{z_2 z_4}+\frac{z_3}{z_4}+\frac{z_2}{z_5}+\frac{1}{z_5}+\frac{z_2 z_4}{z_3 z_1}+\frac{z_5}{z_1}+\frac{z_2}{z_4 z_1}+\frac{z_3}{z_5 z_1} \,, \label{dieBpsi}
\end{eqnarray}
where $z_1, \ldots, z_5$ represent the Cartan generators, which one has to integrate on within the Molien-Weyl formula.
Given the above expression of the characters, we compute the plethystic exponentials
\begin{eqnarray}
\label{10dB}
PE[\chi_V u] = \prod_{i=1}^{10} (1 - \chi_{V, i} u) \,, ~~~~~~
PE[\chi_\psi t] = \prod_{i=1}^{16} \frac{1}{(1 - \chi_{\psi, i} t)} \,, ~~~~~~
\end{eqnarray}
where $\chi_{V, i}$ and $\chi_{\psi, i}$ are the $i$-th components of the characters \eqref{dieB} and \eqref{dieBpsi}, respectively.  \\
In order to implement the Molien-Weyl formula, we also need the invariant measure associated with $SO(10)$, which is given by
\begin{eqnarray}
\label{dieA}
\diff \mu_{SO(10)} &=& \frac{1}{z_1^7 z_2^7 z_3^7 z_4^7 z_5^7}
\left(z_1^2-z_2\right) \left(z_2-1\right) \left(z_1-z_3\right) \left(z_1 z_2-z_3\right) \left(z_1 z_3-z_2\right) \left(z_1 z_3-z_2^2\right) \left(z_3-z_4^2\right) \nonumber \\
&\times &
\left(z_1 z_4-z_5\right) \left(z_1 z_5-z_4\right) \left(z_1 z_5-z_2 z_4\right) \left(z_1 z_4-z_2 z_5\right) \left(z_2 z_5-z_3 z_4\right) \left(z_2 z_4-z_3 z_5\right)\nonumber \\
&\times &
 \left(z_2-z_4 z_5\right) \left(z_1 z_3-z_4 z_5\right) \left(z_1 z_4 z_5-z_3\right) \left(z_1 z_4 z_5-z_2 z_3\right) \left(z_1 z_3-z_2 z_4 z_5\right)
 \nonumber \\
&\times &
  \left(z_2 z_4 z_5-z_3^2\right) \left(z_3-z_5^2\right) \frac{1}{(2\pi i)^5} \diff z_1 \diff z_2 \diff z_3 \diff z_4 \diff z_5 \,.
\end{eqnarray}
The overall factor $\frac{1}{z_1^7 z_2^7 z_3^7 z_4^7 z_5^7}$ in the above expression contains also the integration measure on the Cartan subalgebra. \\
Now, considering $\mathfrak{g}/\mathfrak{h}=\mathfrak{siso}(1|10)/ \mathfrak{so}(10)$ and $\mathfrak{h}'=\mathfrak{h}=\mathfrak{so}(10)$, one can compute the Hilbert-Poincaré polynomial, which results to be 
\begin{align}
\label{dieC}
P(\Omega^{\bullet}(\mathfrak{g}/\mathfrak{h}, \mathfrak{h}' ; \mathbb C ),t,u) &=
1-t^2 \left(u+u^5+u^9\right)+u^{10}
\,.
\end{align}
Then, implementing the MC equations \eqref{die0A}, the polynomial \eqref{dieC} boils down to
\begin{eqnarray}
\label{dieCA}
P(\Omega^{\bullet}(\mathfrak{g}/\mathfrak{h}, \mathfrak{h}' ; \mathbb C ),t,t^2) = 1 - t^4 - t^{12} \,,
\end{eqnarray}
which has the following interpretation in terms of cocycles: 
\begin{align}\label{dieD}
-t^4:\qquad  & \omega^{(3)} = {\frac{i}{2}} \bar{\psi} \gamma_{a} \psi  \, V^a \,, ~~~~~\nonumber \\
- t^{12}:\qquad & \omega^{(7)} = {\frac{i}{5!}} \bar{\psi} \gamma_{a_1\ldots a_5} \psi V^{a_1} \wedge \cdots \wedge  V^{a_5} \,, 
\end{align}
where $\gamma_{a_1\ldots a_5}  = \gamma_{[a_1} \ldots \gamma_{a_5]}$. 
The closure of the two invariants above relies in the following $4\psi$ Fierz identities:
\begin{eqnarray}
\label{dieE}
\bar{\psi} \gamma^a \psi \bar{\psi} \gamma_a \psi =0\,, ~~~~~
\bar{\psi} \gamma^a \psi \bar{\psi} \gamma_{a b c d e} \psi =0 \,.
\end{eqnarray}
In order to build the FDA, we now have to add new forms to compensate the two cocycles in \eqref{dieD}, that is
\begin{eqnarray}
\label{dieF}
\diff B^{(2)} = \omega^{(3)}\,, ~~~~~~~
 \diff B^{(6)} = \omega^{(7)}\,, 
\end{eqnarray}
with scales $t^4$ and $t^{12}$, respectively. The form $B^{(2)}$ is interpreted as the well-known Kalb-Ramond 2-form of $N=1$, $D=10$ supergravity, which couples to the fundamental string F1. Indeed, the cocycle $\omega^{(3)}$ corresponds to the usual Wess-Zumino term. On the other hand, the cocycle $\omega^{(7)}$ does not have a straightforward interpretation. However, let us mention that $N=1$, $D=10$ supegravity 
is an anomalous theory, since there is only a chiral gravitino $\psi^\alpha$, therefore the 6-form $B^{(6)}$ couples to a six-dimensional worldvolume (M5-brane) which might correspond to the anomaly inflow. However, we can at least say that $\omega^{(7)}$ is the Hodge-dual on spacetime of the $\omega^{(3)}$ form, associated with the magnetic dual of F1. \\
Taking \eqref{dieF} into account, the polynomial \eqref{dieCA} is modified as follows: 
\begin{eqnarray}
\label{dieG}
P_{\text{FDA}}(\Omega^{\bullet}(\mathfrak{g}/\mathfrak{h}, \mathfrak{h}' ; \mathbb C ),t,t^2) = \frac{1 - t^4 - t^{12}}{(1-t^4)(1-t^{12})} = 
1 - t^{16} + O\left(t^{20}\right) \,.
\end{eqnarray}
In particular, we have the cocycle\footnote{{The class corresponding to $t^{16}$ actually emerges from the trivialization of one of the two classes previously discussed. In fact, for example, $\omega^{(9)} = B^{(2)} \wedge \omega^{(7)}$ is already closed and nonexact. What happens when we introduce both trivializers is that the two nonexact closed forms $B^{(2)} \wedge \omega^{(7)}$ and $B^{(6)} \wedge \omega^{(3)}$ differ by an exact term.}}
\begin{eqnarray}
\label{dieH}
\omega^{(9)} = \frac12 \left( B^{(2)}\wedge \omega^{(7)} - \omega^{(3)} \wedge B^{(6)} \right)\,.
\end{eqnarray}
The latter can then be removed by the inclusion of a new 8-form $B^{(8)}$, which scales as $t^{16}$. One can then prove that there are still new cocycles at the next level and that, actually, one is left with an infinite series, which means giving rise, in principle, to an infinite series of higher-degree differential forms.

\subsection{$N=2$ type IIA flat rigid superspace}

We will now move on to the application of the Molien-Weyl formula to the $D=10$, $N=2$ type IIA flat rigid superspace case. We therefore consider the coset superspace $G/H={((U(1)_L \times U(1)_R)\ltimes sISO(2|10))/(SO(10)\times U(1)_L \times U(1)_R))}$, whose associated supervielbein satisfies
\begin{eqnarray}
\label{IIA-B}
\nabla V^a = {\frac{i}{2}} \left( \bar{\psi}_L \gamma^a \psi_L + \bar{\psi}_R \gamma^a \psi_R \right) \,, ~~~~~~
\nabla \psi_L =0\,, ~~~~~~
\nabla \psi_R =0\,,
\end{eqnarray}
where ${\psi_{L/R}=\pm \gamma_{11} \psi_{L/R}}$ refer to the two Majorana-Weyl spinors of opposite chirality. \\
The application of the Molien-Weyl formula to the present case yields the following Hilbert series:
\begin{align}\label{IIA-C}
P(\Omega^{\bullet}(\mathfrak{g}/\mathfrak{h}, \mathfrak{h}' ; \mathbb C ),t,u) &= \frac{1-2 t^2 u+\left(t^4+t^2\right) u^2-2 t^4 u^3+\left(t^6+2 t^4+t^2\right) u^4}
{(1-t^2)(1-t^4)}
\nonumber \\ &-\frac{
2\left( t^6+t^4+t^2\right) u^5}{(1-t^2)(1-t^4)}
\nonumber \\ &+ \frac{\left(t^6+2 t^4+t^2\right) u^6-2 t^4 u^7+\left(t^4+t^2\right) u^8-2 t^2 u^9+u^{10}}{(1-t^2)(1-t^4)} \,,
\end{align}
where we have considered {$\mathfrak{g}/\mathfrak{h}=((\mathfrak{u}(1)_L \oplus \mathfrak{u}(1)_R)\ltimes \mathfrak{siso}(2|10))/(\mathfrak{so}(10) \oplus \mathfrak{u}(1)_L \oplus \mathfrak{u}(1)_R)$ and $\mathfrak{h}'=\mathfrak{h}=\mathfrak{so}(10) \oplus \mathfrak{u}(1)_L \oplus \mathfrak{u}(1)_R$}.
When we implement the MC equations \eqref{IIA-B}, the above polynomial reduces to
\begin{eqnarray}\label{IIA-D}
P(\Omega^{\bullet}(\mathfrak{g}/\mathfrak{h}, \mathfrak{h}' ; \mathbb C ),t,t^2) = \frac{1-t^4 +t^6}{1-t^2} \,.
\end{eqnarray}
Let us now analyze the cocycles. The factor $1/(1-t^2)$ is given in terms of a commuting 2-form which is compensated by a 1-form $C^{(1)}$,\footnote{{Let us mention that in applying our method we do not see the 0-form dilaton field actually appearing in the theory.}}
\begin{eqnarray}
\label{IIA-DA}
\omega^{(2)} = - \bar{\psi}_R \psi_L = \diff C^{(1)} \,,
\end{eqnarray}
where $C^{(1)}$ is interpreted as the Ramond-Ramond potential which couples to the D0-brane. \\
{Note that the term $t^6$ at numerator in \eqref{IIA-D} arises from cancellations in the series expansion of \eqref{IIA-C}. It corresponds to a $(t^2)^3$ contribution, that is $\omega^{(2)}\wedge \omega^{(2)}\wedge\omega^{(2)} $.} \\
By including $C^{(1)}$, the Hilbert-Poincar\'e series gets modified as follows: 
\begin{eqnarray}
\label{IIA-E}
P_{\text{FDA}}(\Omega^{\bullet}(\mathfrak{g}/\mathfrak{h}, \mathfrak{h}' ; \mathbb C ),t,t^2) = \frac{1-t^4 +t^6}{1-t^2} (1- t^2) = 1 - t^4 + t^6 \,.
\end{eqnarray}
This polynomial leads to an infinite-dimensional FDA: let us consider the first steps in the trivialization of the Hilbert series. \\
We start by compensating the $-t^4$ term, which corresponds to the 3-form cocycle 
\begin{eqnarray}
\label{IIA-F}
\omega^{(3)} = - i \left(\bar{\psi}_L \gamma_a \psi_L - \bar{\psi}_R \gamma_a \psi_R \right) V^a \,,
\end{eqnarray}
by means of a 2-form $B^{(2)}$, that is
\begin{equation}
     \omega^{(3)} = \diff B^{(2)} \,.
\end{equation}
Such 2-form corresponds to the Kalb-Ramond form of $N=2$ supergravity and it couples to the F1-string \cite{Johnson}. The inclusion of $B^{(2)}$ further modifies the FDA series as follows: 
\begin{eqnarray}
\label{IIA-G}
P'_{\text{FDA}}(\Omega^{\bullet}(\mathfrak{g}/\mathfrak{h}, \mathfrak{h}' ; \mathbb C ),t,t^2) = \frac{1 - t^4 + t^6}{1-t^4} = 1+t^6+t^{10}+O\left(t^{13}\right) \,,
\end{eqnarray}
where new cocycles appear. In particular, we have 
\begin{eqnarray}
\label{IIA-GA}
    \omega^{(4)} = {- \omega^{(3)} \wedge C^{(1)}} + \frac{1}{2} \left( \bar{\psi}_L \gamma_{ab} \psi_R + \bar{\psi}_R \gamma_{ab} \psi_L \right) V^a V^b \,,
\end{eqnarray}
which can be compensated by a 3-form potential $C^{(3)}$,
\begin{equation}
    \omega^{(4)} = \diff C^{(3)} \,.
\end{equation}
The $C^{(3)}$ field is precisely the Ramond-Ramond charge appearing in the type IIA supergravity spectrum.\\
Let us mention here, that $\omega^{(4)}$ can be seen as emerging from the dimensional reduction of the 4-form cocycle of the $D=11$ case that will be derived in Sec. \eqref{sec6}, whereas $C^{(3)}$ arises from the dimensional reduction of the 3-form gauge potential appearing in $D=11$ supergravity.\\ {We have therefore recovered the higher-form gauge potentials of type IIA supergravity, namely $C^{(1)}$, $B^{(2)}$, and $C^{(3)}$ (see, e.g., \cite{DAuria:2008hmx}).} \\
With the inclusion of $C^{(3)}$, the series becomes 
\begin{eqnarray}
\label{IIA-G0}
P''_{\text{FDA}}(\Omega^{\bullet}(\mathfrak{g}/\mathfrak{h}, \mathfrak{h}' ; \mathbb C ),t,t^2) = \frac{1 - t^4 + t^6}{1-t^4} (1-t^6) = 1+t^{10}-t^{12}+O\left(t^{13}\right) \,,
\end{eqnarray}
producing an infinite number of cocycles and associated trivializing forms.

\subsection{$N=2$ type IIB flat rigid superspace}

In this section, we apply the Molien-Weyl integral method to the $D=10$, $N=2$ type IIB flat rigid superspace case, computing Lorentz-invariant cocycles. We consider the coset superspace $G/H={(SO(2)\ltimes sISO(2|10))/(SO(10)\times SO(2))}$. In this case, the MC equations are
\begin{eqnarray}
\label{IIB-B}
\nabla V^a = {\frac{i}{2}} \bar{\psi}^A \gamma^a \psi_A  \,, ~~~~~~
\nabla \psi_A =0\,,
\end{eqnarray}
where $A=1,2$ is the {$SO(2)$} $R$-symmetry index. \\
The result of the Molien-Weyl computation (with $\mathfrak{g}/\mathfrak{h}=(\mathfrak{so}(2)\ltimes \mathfrak{siso}(2|10))/(\mathfrak{so}(10)\oplus \mathfrak{so}(2))$ and {$\mathfrak{h}'=\mathfrak{h}=\mathfrak{so}(10)$}) is 
\begin{align}\label{IIB-C0}
P(\Omega^{\bullet}(\mathfrak{g}/\mathfrak{h}, \mathfrak{h}' ; \mathbb C ),t,u,w) &= \frac{1}{1-t^4}
\left( 1 
- u t^2  \left(w^2+\frac{1}{w^2}+1\right)
+u^2 t^4  \left(w^2+\frac{1}{w^2}+1\right)
-u^3 \left(t^2+ t^6\right) 
\right. \nonumber \\ 
&+ \left.  u^4 t^4  \left(w^2+\frac{1}{w^2}+2 \right)
-u^5 \left(t^2+ t^6 \right) \left( w^2+\frac{1}{w^2} {+1} \right)
+u^6 t^4 \left( w^2+\frac{1}{w^2}+2 \right)
\right. \nonumber \\ 
&- \left. 
u^7 \left(t^2+ t^6\right) 
+ u^8 t^4 \left(w^2+\frac{1}{w^2}+1\right)
- u^9 t^2  \left( w^2+ \frac{1}{w^2}+1\right)
+u^{10} \right)
\end{align}
where the parameter $w$ is related to {$SO(2)$}. Let us give the explicit form of some of the terms appearing in the above expression:
\begin{align}\label{IIB-CB}
\frac{1}{1-t^4} : \qquad & \omega^{(4)} = \bar{\psi}^A \gamma^a \psi^B \bar{\psi}_A \gamma^a \psi_B \,, \nonumber \\
- u t^2 \left(w^2+\frac{1}{w^2}+1\right) : \qquad & \omega^{(3)|AB} = {\frac{i}{2}} V^a  \bar{\psi}^{(A} \gamma_a \psi^{B)} {= {\frac{i}{2}} V^a  \bar{\psi}^{(A} \gamma_a \psi^{B)_0} + \frac{i}{4} \delta^{AB} V^a \bar{\psi}^C \gamma_a \psi_C} \,, \nonumber \\
+u^2 t^4 \left(w^2+\frac{1}{w^2}+1\right) : \qquad &  \omega^{(6)|AB} = V^a V^b \epsilon_{CD} \bar{\psi}^C \gamma_{abc} \psi^D  \bar{\psi}^{(A} \gamma^c \psi^{B)} \,, \nonumber \\
- t^2 u^3 : \qquad & \omega^{(5)} = \epsilon_{AB}  \bar{\psi}^A \gamma_{abc} \psi^B   V^a V^b V^c \,, \nonumber \\
- t^6 u^3 : \qquad & \omega^{(9)} =\delta_{AD} \epsilon_{BC}  \delta_{EF}
\bar{\psi}^A \gamma^a \psi^B  \bar{\psi}^C \gamma^b \psi^D
\bar{\psi}^E \gamma_{abcde} \psi^F   V^c V^d V^e \,, \nonumber \\
u^{10} : \qquad & \omega^{(10)} =\epsilon_{a_1 \ldots a_{10}} V^{a_1} \wedge \cdots \wedge V^{a_{10}}\,,
\end{align}
where with $(AB)_0$ we denote tracelessness in the symmetric indices $(AB)$.
Implementing the MC equations \eqref{IIB-B}, the Hilbert-Poincaré polynomial \eqref{IIB-C0} boils down to 
\begin{eqnarray}\label{IIB-C}
P(\Omega^{\bullet}(\mathfrak{g}/\mathfrak{h}, \mathfrak{h}' ; \mathbb C ),t,t^2,w) = {1 - t^4 \left(w^2 + \frac{1}{w^2}\right)} \,.
\end{eqnarray}
The second term in the above expression indicates the presence of a $3$-cocycle in the symmetric traceless representation of $SO(2)$, with two independent components. {The explicit form of this object is
\begin{equation}
    \omega^{(3)|(AB)_0} = \frac{i}{2} V^a \bar{\psi}^{(A} \gamma_a \psi^{B)_0} \,,
\end{equation}
which can by trivialized by introducing two corresponding $2$-forms 
\begin{align}\label{blah}
    \diff B^{(2)|(AB)_0}=\omega^{(3)|(AB)_0}.
\end{align}
These actually correspond to the Kalb-Ramond and Ramond-Ramond $2$-forms appearing in the type IIB supergravity spectrum. To compare with the relevant literature, see for example \cite{Castellani:1987pi,Castellani:1993ye}, it is sufficient to rearrange the independent components in a representation of $U(1)$ and to multiply the resulting FDA structure by the coset representatives of $SU(1,1)/SO(2)$, describing the dilaton and the Ramond-Ramond $0$-form.}
{In the vacuum, where the scalar fields are constant, this ``dressing'' can automatically be avoided.}\\
The polynomial of the associated FDA is
\begin{eqnarray}
\label{IIB-D}
P_{\text{FDA}}(\Omega^{\bullet}(\mathfrak{g}/\mathfrak{h}, \mathfrak{h}' ; \mathbb C ),t,t^2,w) = \frac{1 - t^4 \left(w^2 + \frac{1}{w^2}\right)}{
(1 - t^4 w^2)(1 - \frac{t^4}{w^2})} = 1-t^8+O\left(t^{12}\right) \,,
\end{eqnarray}
leading to an additional cocycle in the trivial $SO(2)$ representation. This corresponds to a $5$-form $\omega^{(5)}$
which can be compensated by the inclusion of a 4-form gauge potential $C^{(4)}$. {The latter shall correspond to the Ramond-Ramond $4$-form with selfdual field strength.} Then, due to the inclusion of the 4-form $C^{(4)}$, the Hilbert series {becomes 
\begin{align}
P'_{\text{FDA}}(\Omega^{\bullet}(\mathfrak{g}/\mathfrak{h}, \mathfrak{h}' ; \mathbb C ),t,t^2,w) &= \frac{1 - t^4 \left(w^2 + \frac{1}{w^2}\right)}{
(1 - t^4 w^2)(1 - \frac{t^4}{w^2})(1-t^8)} \nonumber \\
& = 1- t^{12} \left(w^2+\frac{1}{w^2}\right) - t^{16} \left( w^4 + \frac{1}{w^4} +1 \right) + O \left( t^{20} \right) \,,
\end{align}
signalling the presence of an odd $7$-form cocycle $\omega^{(7)|(AB)_0}$ in the symmetric traceless representation of $SO(2)$ (corresponding to the Hodge-dual on spacetime of the $3$-form introduced above), together with an odd singlet $9$-form cocycle $\omega^{(9)}$ in the trivial $SO(2)$ representation and other two odd $9$-form cocycles $\omega^{(9)|(AB)_0 (CD)_0}$ at order $t^{16}$, which form a doublet giving rise to the dual description of the scalar fields of type IIB supergravity. Let us mention here that $\omega^{(7)|(AB)_0}$ may be seen as corresponding to a doublet of $SU(1,1)/SO(2)$ and that the $6$-form gauge potentials $B^{(6)|(AB)_0}$ trivializing it couple to the D5-brane/NS5-brane. On the other hand, the trivializer of the singlet $9$-form cocycle $\omega^{(9)}$ is a $8$-form $B^{(8)}$ which couples to the D7-brane (see, e.g., \cite{Johnson,Polchinski1,Polchinski2}).}

\section{$D=11$ spacetime dimensions}\label{sec6}

In this section, we apply the Molien-Weyl formula to the $D=11$, $N=1$ superspace case, showing that, starting from the analysis of the coset space $G/H=sISO(1|11)/SO(11)$, one can reproduce the whole FDA describing the vacuum structure underlying $D=11$ supergravity \cite{Cremmer:1978km,DFd11} (see also \cite{CDF}). \\
The coset space can be dually described in terms of the supervielbein $(V^a,\psi^\alpha)$, satisfying the MC equations\footnote{Here, $a,b,\ldots=0,1,\ldots,10$ are vector indices, $\alpha=1,\ldots,32$ a spinorial index, and $\psi^\alpha$ is a Majorana gravitino. For simplicity, in the following we will frequently omit writing the spinorial index.}
\begin{eqnarray}\label{MC11d}
{\diff} V^a =  {\frac{i}{2}} \bar{\psi} \Gamma^a \psi \,,
~~~~~
{\diff} \psi =0 \,,
\end{eqnarray}
where ${\diff}$ is the nilpotent differential operator and the capital Greek letter $\Gamma$ denote the $D=11$ Dirac matrices. \\
We will now compute all possible Lorentz-invariant combinations of $V^a, \psi^\alpha$ with the Molien-Weyl formula.
To this end, we recall that $SO(11)$ has rank $5$ and that its Dynkin label is $B_5$.\\
Two key ingredients needed for this analysis are the character $\chi_{V}(z_1, \ldots, z_5)$ of the 11-dimensional vectorial representation of $V^a$ and the character $\chi_{\psi}(z_1, \ldots, z_5)$ of the 32-dimensional spinorial representation of $\psi^\alpha$, namely
\begin{eqnarray}
\label{11dAa}
\chi_{V}(z_1, \ldots, z_5) &=&1 + \frac{z_5^2}{z_4}+z_1+\frac{z_2}{z_1}+\frac{z_3}{z_2}+\frac{z_4}{z_3}+\frac{1}{z_1}+\frac{z_1}{z_2}+\frac{z_2}{z_3}+\frac{z_3}{z_4}+\frac{z_4}{z_5^2} \,, \\
\chi_{\psi}(z_1, \ldots, z_5)&=& \frac{z_5 z_1}{z_2}+\frac{z_5 z_1}{z_3}+\frac{z_3 z_5 z_1}{z_2 z_4}+\frac{z_5 z_1}{z_4}+\frac{z_3 z_1}{z_2 z_5}+\frac{z_4 z_1}{z_2 z_5}+\frac{z_4 z_1}{z_3 z_5}+\frac{z_1}{z_5} 
\nonumber \\ &+&\frac{z_5}{z_2}+\frac{z_2 z_5}{z_3}+\frac{z_5}{z_3}+\frac{z_2 z_5}{z_4}+\frac{z_3 z_5}{z_2 z_4}+\frac{z_3 z_5}{z_4}+\frac{z_5}{z_4}+z_5+\frac{z_2}{z_5}+\frac{z_3}{z_2 z_5}
\nonumber \\ &+&
\frac{z_3}{z_5}+\frac{z_4}{z_2 z_5}+\frac{z_2 z_4}{z_3 z_5}+\frac{z_4}{z_3 z_5}+\frac{z_4}{z_5}
+\frac{1}{z_5}+\frac{z_2 z_5}{z_3 z_1}+\frac{z_2 z_5}{z_4 z_1}+\frac{z_3 z_5}{z_4 z_1}
\nonumber \\ &+&
\frac{z_5}{z_1}+\frac{z_2}{z_5 z_1}+\frac{z_3}{z_5 z_1}+\frac{z_2 z_4}{z_3 z_5 z_1}+\frac{z_4}{z_5 z_1} \,. \label{11dAb}
\end{eqnarray}
From those characters one can compute the plethystic exponentials 
\begin{eqnarray}
\label{11dB}
PE[\chi_{V} u] = \prod_{i=1}^{11} (1 - \chi_{V, i} u) \,, ~~~~~~
PE[\chi_{\psi} t] = \prod_{i=1}^{32} \frac{1}{(1 - \chi_{\psi, i} t)} \,, ~~~~~~
\end{eqnarray}
where $\chi_{V, i}$ and $\chi_{\psi, i}$ are the $i$-th components of the characters \eqref{11dAa} and \eqref{11dAb}, respectively. \\
To implement the Molien-Weyl formula, we also need the invariant measure associated with $SO(11)$. We get
\begin{align}
\label{11dC}
\diff \mu_{SO(11)}& = \frac{1}{z_1^8 z_2^8 z_3^8 z_4^8 z_5^9}
\left(1-z_1\right) \left(1-z_2\right) \left(z_1-z_2\right) \left(z_1^2-z_2\right) \left(z_1-z_3\right) \left(z_2-z_3\right) \left(z_1 z_2-z_3\right) \left(z_2-z_1 z_3\right)  \nonumber \\
&\times 
\left(z_2^2-z_1 z_3\right) \left(z_2-z_4\right) \left(z_3-z_4\right) \left(z_1 z_3-z_4\right) \left(z_3-z_1 z_4\right) \left(z_2 z_3-z_1 z_4\right) \left(z_1 z_3-z_2 z_4\right) 
 \nonumber \\
&\times 
\left(z_3^2-z_2 z_4\right) \left(z_3-z_5^2\right) \left(z_4-z_5^2\right) \left(z_1 z_4-z_5^2\right) \left(z_4-z_1 z_5^2\right) \left(z_2 z_4-z_1 z_5^2\right) \left(z_1 z_4-z_2 z_5^2\right) 
 \nonumber \\
&\times 
\left(z_3 z_4-z_2 z_5^2\right) \left(z_2 z_4-z_3 z_5^2\right) \left(z_4^2-z_3 z_5^2\right) {\frac{1}{(2\pi i)^5} \diff z_1 \diff z_2 \diff z_3 \diff z_4 \diff z_5} \,,
\end{align}
where the overall factor $\frac{1}{z_1^8 z_2^8 z_3^8 z_4^8 z_5^9}$ also contains the integration measure on the Cartan subalgebra.\\ 
The Hilbert-Poincaré series with
$\mathfrak{g}/\mathfrak{h}=\mathfrak{siso}(1|11)/ \mathfrak{so}(11)$ and $\mathfrak{h}'=\mathfrak{h}=\mathfrak{so}(11)$ is in the form
\begin{eqnarray}
\label{11dD}
P(\Omega^{\bullet}(\mathfrak{g}/\mathfrak{h}, \mathfrak{h}' ; \mathbb C ),t,u) =\oint_{|z_i|=1, i=1,\ldots,5} PE[\chi_{V}u] PE[ \chi_{\psi}t] \diff\mu_{SO(11)} \,,
\end{eqnarray}
which, after the residue integration, gives
\begin{align}
\label{11dE}
P(\Omega^{\bullet}(\mathfrak{g}/\mathfrak{h}, \mathfrak{h}' ; \mathbb C ),t,u)& = \frac{1}{1- t^4}
\Big(1 + t^4 u^8-t^4 u^7+t^4 u^4-t^4 u^3+t^2 u^{10}-t^2 u^9+t^2 u^2-t^2 u \nonumber \\
& -\left(-t^6-t^4-t^2\right) u^6-\left(t^6+t^4+t^2\right) u^5-u^{11}\Big) \,.
\end{align}
Let us now give an interpretation to some of the above terms:
\begin{align}\label{11dF}
\frac{1}{1-t^4} : \qquad & \omega^{(4)}_1 = \bar\psi \Gamma_{abcde} \psi  \bar\psi \Gamma^{abcde} \psi \,, \nonumber \\
-t^2 u: \qquad & \omega^{(3)} = {\frac{i}{2}} \bar\psi \Gamma_{a} \psi V^a \,, \nonumber \\
t^2 u^2 : \qquad & \omega^{(4)}_2 ={\frac{1}{2}} \bar\psi \Gamma_{ab} \psi V^a V^b   \,, \nonumber \\
-t^4 u^3 : \qquad & \omega^{(3)} \wedge \omega^{(4)}_2 = {\frac{i}{4}} \bar\psi \Gamma_{[a} \psi   \bar\psi \Gamma_{bc]} \psi  V^a V^b V^c \,, \nonumber \\
u^{11}: \qquad & \omega^{(11)} = \epsilon_{a_1 \ldots a_{11}} V^{a_1} \wedge \cdots \wedge V^{a_{11}} \,.
\end{align}
The factor $1/(1-t^4)$ stands for the powers of the invariant expression 
$\bar\psi \Gamma_{abcde} \psi  \bar\psi \Gamma^{abcde} \psi$, which is commuting and therefore it can appear with any power. 
{Notice that, in principle, one could consider other two expressions at the same scale level: $\bar\psi\Gamma_a\psi \bar\psi \Gamma^a\psi$ and 
$\bar\psi\Gamma_{ab}\psi \bar\psi \Gamma^{ab}\psi$. However, these expressions are not independent, precisely because of the following $3\psi$ relation:
\begin{equation}
    A \, \Gamma_a\psi \bar\psi \Gamma^a\psi + B \, \Gamma_{ab}\psi \bar\psi \Gamma^{ab}\psi + C \, \Gamma_{abcde} \psi  \bar\psi \Gamma^{abcde} \psi =0 \,, \qquad A - 10 \, B - 6 \cdot 5!\, C = 0 \,.
\end{equation}
In particular, the latter yields the $3\psi$ Fierz identities
\begin{align}\label{usefulfierz11d}
    & 5 \Gamma_a\psi \bar\psi \Gamma^a\psi + \frac{1}{2} \Gamma_{ab}\psi \bar\psi \Gamma^{ab}\psi = 0 \,, \nonumber \\
    & 6 \Gamma_a\psi \bar\psi \Gamma^a\psi + \frac{1}{5!} \Gamma_{abcde} \psi  \bar\psi \Gamma^{abcde} \psi = 0 \,,
\end{align}
showing that we are left with a single independent $4\psi$ invariant.} \\
Then, implementing the MC equations, we finally get
\begin{eqnarray}
\label{11dG0}
P(\Omega^{\bullet}(\mathfrak{g}/\mathfrak{h}, \mathfrak{h}' ; \mathbb C ),t,t^2) = 1 + t^6 \,,
\end{eqnarray}
where the $t^6$ stands for $\omega^{(4)}_2 = {\frac{1}{2}} \bar\psi \Gamma_{ab} \psi V^a V^b$. Thus, as it is well-known (see, e.g., \cite{CDF,DFd11}), in the present case the collection of invariants reduces to one cohomology class only. \\ 
In order to construct the FDA, we have to add a 3-form $A^{(3)}$ scaling with $t^6$, in such a way that 
\begin{eqnarray}
\label{11dH}
\diff A^{(3)} = \omega^{(4)}_2 \,.
\end{eqnarray}
Then, the resulting polynomial becomes 
\begin{eqnarray}
\label{11dG1}
 P_{\text{FDA}}(\Omega^{\bullet}(\mathfrak{g}/\mathfrak{h}, \mathfrak{h}' ; \mathbb C ),t,t^2) =  (1 + t^6)(1 - t^6) = 1 - t^{12} \,.
\end{eqnarray} 
This means that the FDA is not complete: indeed, one can immediately see, that there is a new cohomology class
\begin{eqnarray}
\label{11dL}
\omega^{(7)} = {15} A^{(3)} \wedge \omega^{(4)}_2 {+ \frac{i}{2}} \bar \psi \Gamma_{a_1 \ldots a_5} \psi V^{a_1} \wedge \cdots \wedge V^{a_5} \,,
\end{eqnarray}
whose closure in fact relies on the $3\psi$ Fierz identity
\begin{equation}\label{usefulfierz}
    \Gamma_{[a_1a_2} \psi \bar{\psi} \Gamma_{a_3a_4]} \psi + \frac{1}{3} \Gamma_{a_1\ldots a_5} \psi \bar{\psi} \Gamma^{a_5} \psi = 0 \,.
\end{equation}
To cancel the class \eqref{11dL}, one needs to introduce a further commuting potential $B^{(6)}$ such that 
$\diff B^{(6)} = \omega^{(7)}$. Therefore, the final expression for the Hilbert-Poincar\'e series is 
\begin{eqnarray}
\label{11dG2}
P_{\text{FDA}}(\Omega^{\bullet}(\mathfrak{g}/\mathfrak{h}, \mathfrak{h}' ; \mathbb C ),t,t^2) =  \frac{(1 + t^6)(1 - t^6)}{1- t^{12}} = \frac{1- t^{12}}{1- t^{12}} = 1 \,.
\end{eqnarray} 
Therefore, we have recovered the full FDA describing the vacuum structure underlying $D=11$ supergravity by using the Molien-Weyl integral method.

\section{$D=12$ spacetime dimensions}\label{sec7}

In this section, we focus on the computation of invariants and cocycles for $12$-dimensional theories. This number of dimensions is needed for any hypothetical nonperturbative completion of type IIB superstrings.\\ 
However, the standard signature $(11,1)$ would produce, through dimensional reduction, higher spins in the $4$-dimensional theory. As it is known, a finite number of higher-spin particles are problematic from a QFT point of view. A solution to this problem is achieved by instead considering a $(10,2)$ signature. In this case, Majorana-Weyl spinors exist and the compactified theory can be argued to be well-behaved.
The low energy description of such a hypothetical ``Father'' theory, F-theory, is 12-dimensional supergravity. In the flat case, the construction of an action principle for such a theory has been proven to be challenging both in the component approach and with the known superspace techniques \cite{Berends:1979kg,Hewson:1996yh,Nishino:1997gq}. 
There is a twelve dimensional theory using the MacDowell-Mansouri-like construction for $N = 1$ supergravity in $10+2$ dimensions presented in \cite{Castellani:2017vbi}. 
In this framework, we hope that the construction of invariants and cocycles, which as we shall see exactly coincide in this case, will help on this matter.

In this case, the Lorentz symmetry is described by $SO(12)$ and we are dealing with the coset space $G/H={sISO(1|12)}/SO(12)$. 
This superspace can be dually described in terms of the supervielbein $(V^a,\psi)$, where $a,b,\ldots=0,1,\ldots,11$ and $\psi$ is a {Majorana-Weyl} spinor, satisfying the following MC equations {(due, in particular, to the properties of the gamma matrices in $D=12$)}:
\begin{equation}\label{MC12d}
    {\diff} V^a = 0 \,, \quad {\diff} \psi = 0 \,,
\end{equation}
where ${\diff}$ is nilpotent.
We write the vectorial representation of the vielbein $V^a$, 
\begin{equation}
\label{twB}
\chi_V(z_1, \dots, z_6) =  \frac{z_1}{z_2}+z_1+\frac{z_3}{z_2}+\frac{z_4}{z_3}+\frac{z_5 z_6}{z_4}+\frac{z_6}{z_5}+\frac{z_2}{z_3}+\frac{z_3}{z_4}+\frac{z_5}{z_6}+\frac{z_4}{z_5 z_6}+\frac{z_2}{z_1}+\frac{1}{z_1} \,,
\end{equation}
and the spinorial representation of $\psi$, 
\begin{align}
\label{twC}
\chi_\psi(z_1, \dots, z_6) &= \frac{z_5 z_1}{z_2}+\frac{z_5 z_1}{z_4}+\frac{z_6 z_1}{z_3}+\frac{z_3 z_6 z_1}{z_2 z_4}+\frac{z_4 z_1}{z_2 z_5}+\frac{z_1}{z_5}+\frac{z_3 z_1}{z_2 z_6}+\frac{z_4 z_1}{z_3 z_6}+\frac{z_2 z_5}{z_3}+\frac{z_5}{z_3} 
\nonumber \\
&+\frac{z_3 z_5}{z_2 z_4}+\frac{z_3 z_5}{z_4}+\frac{z_6}{z_2}+\frac{z_2 z_6}{z_4}+\frac{z_6}{z_4}+z_6+\frac{z_3}{z_2 z_5}+\frac{z_3}{z_5}+\frac{z_2 z_4}{z_3 z_5}+\frac{z_4}{z_3 z_5}+\frac{z_2}{z_6} \nonumber \\
&+\frac{z_4}{z_2 z_6}+\frac{z_4}{z_6}+\frac{1}{z_6}+\frac{z_2 z_5}{z_4 z_1}+\frac{z_5}{z_1}+\frac{z_2 z_6}{z_3 z_1}+\frac{z_3 z_6}{z_4 z_1}+\frac{z_2}{z_5 z_1}+\frac{z_4}{z_5 z_1}+\frac{z_3}{z_6 z_1}+\frac{z_2 z_4}{z_3 z_6 z_1} \,,
\end{align}
where $z_1, \dots, z_6$ are the integration variables related to the Cartan's of $SO(12)$ (we recall that $SO(12)$ has rank $6$ and its Dynkin label is $D_6$).
The integration measure is given by
\begin{eqnarray}
\label{twA}
\diff \mu_{SO(12)}&=&  \frac{1}{z_1^9 z_2^9 z_3^9 z_4^9 z_5^9 z_6^9}
\left(1-z_2\right) \left(z_1^2-z_2\right) \left(z_1-z_3\right) \left(z_1 z_2-z_3\right) \left(z_2-z_1 z_3\right) \left(z_2^2-z_1 z_3\right) \nonumber \\
&\times& \left(z_2-z_4\right) \left(z_1 z_3-z_4\right) \left(z_3-z_1 z_4\right) \left(z_2 z_3-z_1 z_4\right) \left(z_1 z_3-z_2 z_4\right) \left(z_3^2-z_2 z_4\right) \left(z_4-z_5^2\right) \nonumber \\
&\times&
\left(z_1 z_5-z_6\right) \left(z_5-z_1 z_6\right) \left(z_2 z_5-z_1 z_6\right) \left(z_1 z_5-z_2 z_6\right) \left(z_3 z_5-z_2 z_6\right) \left(z_2 z_5-z_3 z_6\right) \nonumber \\
&\times&
\left(z_4 z_5-z_3 z_6\right) \left(z_3 z_5-z_4 z_6\right) \left(z_3-z_5 z_6\right) \left(z_1 z_4-z_5 z_6\right) \left(z_4-z_1 z_5 z_6\right) \left(z_2 z_4-z_1 z_5 z_6\right)\nonumber \\
&\times& \left(z_1 z_4-z_2 z_5 z_6\right) \left(z_3 z_4-z_2 z_5 z_6\right) \left(z_2 z_4-z_3 z_5 z_6\right) \left(z_4^2-z_3 z_5 z_6\right) \left(z_4-z_6^2\right) \,.
\end{eqnarray}
Consequently, as usual, we construct the plethystic polynomials using $u$ to parametrize the powers of $V^a$ and $t$ as the parameter for $\psi$.
Given that the MC equations \eqref{MC12d} in the flat case are completely trivial, the result of the Molien-Weyl computation immediately gives us the cocycles. 
Considering $\mathfrak{g}/\mathfrak{h}={\mathfrak{siso}(1|12)}/\mathfrak{so}(12)$ and $\mathfrak{h}'=\mathfrak{h}=\mathfrak{so}(12)$, the final result for the Hilbert-Poincaré series is 
\begin{eqnarray}
\label{twE}
P(\Omega^{\bullet}(\mathfrak{g}/\mathfrak{h}, \mathfrak{h}' ; \mathbb C ),t,u) = \frac{1+ t^2 \left(u^{10}+u^6+u^2\right)+t^4 \left(u^8+u^6+u^4\right)+ t^6 u^6+u^{12}}{1-t^4} \,.
\end{eqnarray}
{Here one could then consider $u = t^2$, even though it is not prescribed by the MC equations.} However, 
since all terms have positive coefficients, there is no cancellation between terms, which is equivalent as saying that the differential is trivial. \\
It is then easy to identify the contributions to \eqref{twE}: for example, the first terms are
\begin{align}
\label{twF}
\frac{1}{1-t^4} : \qquad & \omega^{(4)}_1 = \bar{\psi} \hat{\Gamma}_{ab} \psi  \bar{\psi} \hat{\Gamma}^{ab} \psi \,, \nonumber \\
u^2 t^2 : \qquad & \omega^{(4)}_2 = \bar{\psi} \hat{\Gamma}_{ab} \psi  V^a V^b \,, \nonumber \\
u^6 t^2 : \qquad & \omega^{(8)} = \bar{\psi} \hat{\Gamma}_{a_1 \dots a_6} \psi  V^{a_1} \wedge \cdots \wedge V^{a_6} \,,  
\end{align}
where $\hat{\Gamma}_{ab}$ and $\hat{\Gamma}_{a_1 \ldots a_6}$ are gamma matrices in twelve dimensions. 
At the moment, we do not have a clear interpretations of those cocycles and we leave such research to future endeavours. One possible direction is to compare with the brane scan in higher dimension, as in \cite{Blencowe:1988sk}. 

\section{Deformations and Spencer cohomology}\label{sec8}

In the last years, an interesting approach to classify supergravity solutions emerged  \cite{Santi:2010kb,Santi:2011mc,Figueroa-OFarrill:2015rfh,Figueroa-OFarrill:2015tan,deMedeiros:2016srz,Figueroa-OFarrill:2016khp,deMedeiros:2018ooy,Santi:2019kpx}: in particular, the Killing superalgebras of supergravity backgrounds were put into correspondence with \emph{filtered deformations} of subalgebras of the Poincar\'e superalgebras. For instance, in \cite{Figueroa-OFarrill:2015tan} the authors proved the following theorem for $D=11$ supergravity:
\begin{theorem*}
    The Killing superalgebra of a 11-dimensional supergravity background is a filtered subdeformation of the Poincar\'e superalgebra.
\end{theorem*}
Even though the converse is not proven, it is interesting to classify the subdeformations of Poincar\'e algebras to verify, a posteriori, which correspond to Killing superalgebras of supergravity backgrounds. \\
The tool to classify deformations is the \emph{(generalized) Spencer cohomology}, which we briefly describe here for the case of super-Poincar\'e superalgebras (see, e.g., \cite{Cheng-Kac,Figueroa-OFarrill:2015rfh,Figueroa-OFarrill:2016khp} for a general and exhaustive introduction). Given the Poincar\'e superalgebra $\mathfrak{p}$, we can split it (as a vector space) by means of a $\mathbb{Z}$-grading, compatible with the $\mathbb{Z}_2$ grading, as
\begin{equation}
    \mathfrak{p} = \mathfrak{p}_0 \oplus \mathfrak{p}_{-1} \oplus \mathfrak{p}_{-2} = \mathfrak{so} \left( V \right) \oplus S \oplus V \,,
\end{equation}
where $\mathfrak{so} \left( V \right)$ is the Lorentz subspace, $S$ collects the odd generators and $V$ the (even) generators of the translations. Denoting $\mathfrak{p}_- = \mathfrak{p}_{-1} \oplus \mathfrak{p}_{-2}$ the ideal generated by the supertranslations, we define the Spencer cochains to be Lorentz-invariant cochains of $\Pi \mathfrak{p}_-^*$ (i.e., the cochains of $\mathfrak{p_-}$ relative to $\mathfrak{p}_0$) with values on the whole Poincar\'e superalgebra (that is, we are simply considering Chevalley-Eilenberg cochains with values in the module $\mathfrak{p}$, see, e.g., \cite{Carlo-Antonio-CE2}):
\begin{equation}
    C^{\bullet} \left( \mathfrak{p} , \mathfrak{p}_0 ; \mathfrak{p} \right) \equiv \Omega^\bullet \left( \mathfrak{p}_- , \mathfrak{p} \right) \cong \left( \bigwedge^\bullet \Pi \mathfrak{p}_-^* \right)^{\mathfrak{p}_0} \otimes \mathfrak{p} \,,
\end{equation}
where the action is defined via the adjoint representation. The forms in $\Pi \mathfrak{p}^*$ inherit a $\mathbb{Z}$-grading so that a form $\omega \in \Pi S^*$ has grading $+1$ and a form $\omega \in \Pi V^*$ has grading $+2$ (this corresponds to assigning $V^a \to t^2$ and $\psi^\alpha \to t$ in the flat cases, as described in Sec. \ref{workplan}); this implies that the spaces of cochains can be decomposed according to this grading:
\begin{equation}
    \Omega^\bullet \left( \mathfrak{p}_- , \mathfrak{p} \right) = \bigoplus_{i\in \mathbb{Z}} \Omega^{i,\bullet} \left( \mathfrak{p}_- , \mathfrak{p} \right) \,.
\end{equation}
The differential is defined on forms as usual (i.e., via the MC equations) and on vector fields via the adjoint representation, that is
\begin{equation}\label{diffonfields}
    \left( {\diff} X \right) \left( Y \right) = \left[ X , Y \right] \,,
\end{equation}
where with $\left[ \,, \right]$ we denote the the graded Lie bracket. With this differential at hand, one can lift the cochains to a complex and compute the cohomology. The deformations are encoded in $H^2 \left( \mathfrak{p}_- , \mathfrak{p} \right)$: by denoting $\mathpzc{V}^I=\left\lbrace V^a , \psi^\alpha \right\rbrace$ the generators of $\Pi \mathfrak{p}_-^*$ and with $\mathpzc{X}_{\tilde{I}} = \left\lbrace L_{ab} , X_a , q_\alpha \right\rbrace$ the generators of $\mathfrak{p}$, a form $\omega \in H^2 \left( \mathfrak{p}_- , \mathfrak{p} \right)$ will read as $\omega = \omega^{\tilde{I}}_{IJ} \mathpzc{V}^I \wedge \mathpzc{V}^J \otimes \mathpzc{X}_{\tilde{I}}$; the coefficient $\omega^{\tilde{I}}_{IJ}$ can then be used to deform the Lie brackets Poincar\'e superalgebra.

We will now show how to use the Molien-Weyl formula to compute (or to simplify the computations of) the generalized Spencer cohomology for any degree and for different examples of flat rigid superspaces.

\subsection{Deformations of the $D=4$, $N=1$ Killing superalgebra and Spencer cohomology}

As discussed above, we consider the $V^{\alpha \dot\alpha}, \psi^\alpha, \bar{\psi}^{\dot\alpha}$ forms
satisfying the MC equations \eqref{N1A} and we add the generators $X_{\alpha \dot\alpha}, q_\alpha, \bar q_{\dot\alpha}, 
L_{\alpha \beta}, L_{\dot\alpha\dot \beta}$ of the super-Poincar\'e algebra, which satisfy the equations 
\begin{align}
\label{DEFA}
& {\diff} X_{\alpha\dot\alpha} = 0\,, \nonumber \\
& {\diff} q_\alpha = X_{\alpha \dot\alpha} \bar \psi^{\dot\alpha}\,, 
\qquad {\diff} \bar q_{\dot \alpha} = 
X_{\alpha \dot\alpha} \psi^{\alpha}\,, \nonumber \\
& {\diff} L_{\alpha\beta} = q_{(\alpha} \psi_{\beta)} - X_{(\alpha|\dot\alpha} {V^{\dot\alpha}}_{\beta)} \,, \qquad
{\diff} L_{\dot\alpha\dot\beta} = \bar q_{(\dot\alpha} \bar\psi_{\dot\beta)} - X_{\alpha (\dot\alpha} {V^{\alpha}}_{\dot\beta)} \,.
\end{align}
These equations are derived from \eqref{diffonfields}
and, together with \eqref{N1A}, we have ${\diff}^2 = 0$. \\
As in the previous sections, we shall assign a scale to the MC forms 
$[V] = {u}$, $[\psi]=[\bar\psi]=t$, and, analogously, we will also assign the dual scales  $[X]={u^{-1}}$, $[q]=[\bar q]=t^{-1}$, $[L]=t^0$. As discussed above, the forms under considerations belong to the space {$C^{\bullet} \left( \mathfrak{p} , \mathfrak{p}_0 ; \mathfrak{p} \right)$}. \\
In the following paragraphs, we compute the Lorentz-invariant objects by means of a suitable adaptation of the Molien-Weyl formula to the present case.
Besides the plethystic polynomials for $V^{\alpha\dot\alpha}$, $\psi^\alpha$, $\bar\psi^{\dot\alpha}$, we need to take into account for the transformation properties of the generators of $\mathfrak g$ under the Lorentz symmetry. Therefore, we need the factor 
\begin{eqnarray}
\label{DEFC}
F(z, w) = \frac{1}{u} \chi_X(z, w) - \frac{1}{t} \chi_q(z) - \frac{1}{t} \chi_{\bar q}(w) + \chi_L(z) + \chi_{\bar L}(w)  \,,
\end{eqnarray}
where $\chi_X(z, w) = \chi_V(z, w)$, $\chi_q(z) = \chi_\psi(z)$, and $\chi_{\bar \psi}(w) = \chi_{\bar q}(w)$ are the characters of the generators $X$, $q$, and $\bar q$, respectively (and they coincide with the those of $V$, $\psi$, and $\bar{\psi}$, since they 
transform in the same way) and  
 \begin{eqnarray}
\label{DEFD}
\chi_L(z) = z^2 + \frac{1}{z^2}  + 1\,, ~~~~~~~ 
\chi_{\bar L}(w) = w^2 + \frac{1}{w^2}  + 1\,, 
\end{eqnarray}
which are the $(2,0)$ and $(0,2)$ representations. 
The factors {$1/u$ and $1/t$} in \eqref{DEFC} take into account the scales of $X_{\alpha\dot\alpha}$ and $q_\alpha$, $\bar q_{\dot\alpha}$, while the $L$'s are scaleless.  
Then, we can finally write the formula for the Hilbert series in the presence of deformations,
\begin{eqnarray}
\label{DEFE}
P_{\text{def.}}(\Omega^{\bullet}(\mathfrak{g}/\mathfrak{h}, \mathfrak{h}' ; \mathbb C ),t,u) = \oint_{|z|=1} \oint_{|w|=1} F(z, w) PE[\chi_V u] PE[\chi_\psi t] PE[\chi_{\bar \psi} t] \diff \mu_{SU(2) \times SU(2)} \,,
\end{eqnarray}
where the measure is given in \eqref{newUAA1} and the plethystic polynomials are given in \eqref{pleA}. 
In this formula, we introduce only one factor for each generator $X$, $q$, $\bar q$, $L$, $\bar L$, since we consider only the linear sector with only one power of them. \\
The computation of the expression in \eqref{DEFE} by means of the residues formula gives 
\begin{align}
\label{DEFG}
P_{\text{def.}}(\Omega^{\bullet}(\mathfrak{g}/\mathfrak{h}, \mathfrak{h}' ; \mathbb C ),t,u) & = -2 t^4 u^3+3 t^4 u^2-t^4+2 t^2 u^4+t^2 u^3-4 t^2 u^2+2 t^2 u \nonumber \\
& +\frac{t^2}{u}-2 u^4+2 u^3-u^2+2 u-3 \,,
\end{align}
where we have kept the scales of $V$ and $X$ separated from those of 
$\psi$, $\bar{\psi}$, $q$, and $\bar q$. \\
Let us now list some of the above invariants:
\begin{align}\nonumber\label{DEFH}
    -3: \qquad & \omega^{(1)}_1=X_{\alpha\dot\alpha}V^{\alpha\dot\alpha}, \quad \omega^{(1)}_2=q_\alpha \psi^\alpha, \quad \omega^{(1)}_3=\bar q_{\dot \alpha} \bar\psi^{\dot \alpha}\\ \nonumber
    \frac{t^2}{u}:\qquad & \omega^{(2)}_1=X_{\alpha\dot\alpha}\psi^\alpha\bar\psi^{\dot\alpha}\\
    2u:\qquad & \omega^{(2)}_2=V^{\alpha\dot\alpha}\psi_\alpha\bar q_{\dot\alpha},\quad \omega^{(2)}_3=V^{\alpha\dot\alpha}q_\alpha\bar \psi_{\dot\alpha},
\end{align}
The cohomology is easily obtained by setting $u = t^2$:
\begin{eqnarray}\label{DEFI}
P_{\text{def.}}(\Omega^{\bullet}(\mathfrak{g}/\mathfrak{h}, \mathfrak{h}' ; \mathbb C ),t,t^2) = -2+2 t^2 - 2 t^6 + 2 t^8 \,.
\end{eqnarray}
Notice, for example, that the first addendum $-2$ represents the cohomology among the the scaleless terms in \eqref{DEFH}. Indeed, by acting with ${\diff}$ on the three invariant, we get
\begin{eqnarray}
\label{DEFL}
{\diff}
\left(a  X_{\alpha \dot\alpha} V^{\alpha \dot\alpha}  + b 
q_\alpha \psi^\alpha + c \bar q_{\dot \alpha} \bar\psi^{\dot \alpha} \right) =0 \quad \Rightarrow \quad a + b + c = 0 \,.
\end{eqnarray}
Therefore, from three invariants there emerge just two cohomology classes. {In this case, the Spencer cohomology group at scaling weight $2$, $H^2(\mathfrak{p}_-,\mathfrak{p})$, is generated by the representatives
\begin{align}
    \tilde\omega^{(2)}_1=V^{\alpha\dot\alpha}\psi_\alpha \bar q_{\dot\alpha}-\psi_\alpha\psi_\beta L^{\alpha\beta},\qquad \tilde\omega^{(2)}_2=V^{\alpha\dot\alpha}q_\alpha \bar \psi_{\dot\alpha}-\bar\psi_{\dot\alpha}\bar\psi_{\dot\beta} L^{\dot\alpha\dot\beta}.
\end{align}}
This result is in agreement with the discussion in \cite{deMedeiros:2016srz}.

\subsection{Deformations of the $D=11$, $N=1$ Killing superalgebra and Spencer cohomology}

In this section, we study the deformations of the Killing superalgebra for the $D=11$ supergravity 
background, which was discussed in \cite{Santi:2019kpx}, giving an application of our mathematical tools in computing the generalized Spencer cohomology for the $D=11$ case. Killing superalgebras are generated by 
the Killing spinors and Killing vectors of a given supergravity background and the generalized Spencer cohomology 
\cite{Cheng-Kac} computes their deformations as previously discussed. \\
With respect to Sec. \ref{sec6}, we add the generators of the Killing superalgebra, $X_a, q_\alpha, L_{ab}$, in the 
vectorial, spinorial, and adjoint representations, respectively. For them, we have the following 
equations:
\begin{eqnarray}
\label{elevA}
{\diff} X_a =0\,, ~~~~~~
{\diff} q_\alpha = {-\frac{i}{2}} X_a (\Gamma^a\psi)_\alpha\,, ~~~~~~
{\diff} L_{ab} = X_{[a} V_{b]} - {\frac{1}{2}} \bar q\Gamma_{ab} \psi \,,
\end{eqnarray}
which are complemented by 
the MC equations \eqref{MC11d}. \\
In this case, we have to introduce the following prefactor
\begin{eqnarray}
\label{elevC}
F(z_1, \ldots, z_5) = \frac1u \chi_{X}(z_1, \ldots, z_5) - \frac{1}{t} \chi_{q}(z_1, \ldots, z_5) 
+ \chi_{L}(z_1, \ldots, z_5) 
\end{eqnarray}
in the computation of the Molien-Weyl formula for the Hilbert-Poincaré series, 
\begin{eqnarray}
\label{elevB}
P_{\text{def.}}(\Omega^{\bullet}(\mathfrak{g}/\mathfrak{h}, \mathfrak{h}' ; \mathbb C ),t,u) = \oint_{|z_i|=1, i=1,\ldots,5} F(z_1, \ldots, z_5) 
PE[\chi_V u] PE[- \chi_\psi t] \diff\mu_{SO(11)} \,,
\end{eqnarray}
where $\chi_X(z_1, \ldots, z_5) = \chi_V(z_1, \ldots, z_5)$ and $\chi_q(z_1, \ldots, z_5) = \chi_\psi(z_1, \ldots, z_5)$ are the characters of the generators $X$ and $q$, respectively, while $\chi_{L}(z_1, \ldots, z_5)$ is the character of the adjoint representation of $SO(11)$. The explicit expression reads
\begin{eqnarray}\label{elevD}
P_{\text{def.}}(\Omega^{\bullet}(\mathfrak{g}/\mathfrak{h}, \mathfrak{h}' ; \mathbb C ),t,u) &=&
\frac{1}{\left(1- t^4\right) u}
{(u-1) \left[-t^8 u^7+t^6 \left(-\left(u^9-u^8+u^6\right)\right) \right.}
\nonumber \\
&+&t^4 \left(u^{10}+u^9-u^8+u^7-u^6+2 u^5-2 u^4+u\right)
\nonumber \\
&+&t^2 \left(-2 u^{10}+u^9-u^8+u^7-3 u^6+u^5-u^4+u^3-2 u^2-1\right)
\nonumber \\
&+&{\left. u \left(u^{10}+u^8+u^6+u^4+u^2+u+2\right)\right]} \,,
\end{eqnarray}
which describes all possible invariant expressions constructed in terms of the ingredients discussed above. 
Again, the factor $1/(1-t^4)$ is related to the invariant $\bar\psi\Gamma_{abcde} \psi
\bar\psi\Gamma^{abcde} \psi$. \\
In order to compute the Spencer cohomology, we finally set $u= t^2$, obtaining 
\begin{eqnarray}
\label{elveD}
P_{\text{def.}}(\Omega^{\bullet}(\mathfrak{g}/\mathfrak{h}, \mathfrak{h}' ; \mathbb C ),t,t^2)  = -1 - t^6 + t^8 \,,
\end{eqnarray}
which describes three cohomology classes. 
The interpretation is the following:
\begin{align}
\label{elevE}
-1 : \qquad & \omega^{(1)} = X_a V^a - \bar q_\alpha \psi^\alpha \,, \nonumber \\
- t^6 : \qquad & \omega^{(5)} = \left(X_a V^a - \bar q_\alpha \psi^\alpha\right) \wedge  
(\bar\psi \Gamma_{cd} \psi V^c V^d) \,, \nonumber \\
+ t^8 : \qquad & \omega^{(6)} = {i} \bar\psi \Gamma_{abcde} \psi V^a V^b V^c V^d X^e {+ \frac{3}{2}}
(\bar\psi \Gamma_{ab} \psi V^a V^b) \wedge  (\bar \psi \Gamma_{cd} \psi L^{cd} {+ 10 i \, \bar{q}\Gamma_c \psi V^c} ) \,.
\end{align}
Then, we see that 
\begin{eqnarray}
\label{evelE}
 \omega^{(5)}  = {2} \omega^{(1)} \wedge \omega^{(4)}_2 \,, ~~~~~
  \omega^{(6)} = {i} \bar\psi \Gamma_{abcde} \psi V^a V^b V^c V^d X^e {+ 3}
\omega^{(4)}_2 \wedge  (\bar \psi \Gamma_{cd} \psi L^{cd} {+ 10 i \, \bar{q}\Gamma_c \psi V^c}) \,,
\end{eqnarray}
where the explicit expression of the 4-form $\omega^{(4)}_2$ is, in fact, $\omega^{(4)}_2 = {\frac{1}{2}} \bar\psi \Gamma_{ab} \psi V^a V^b$, as given in Sec. \ref{sec6}. 
Note that 
\begin{align}
\label{evelF}
\diff \omega^{(6)} = & {-2}  \bar\psi \Gamma_{abcde} \psi \bar\psi \Gamma^a \psi V^b V^c V^d X^e \nonumber \\
& {+ 3}
\omega^{(4)}_2 \wedge \left[  \bar \psi \Gamma_{cd} \psi \left( X^c V^d - {\frac{1}{2}} \bar q \Gamma^{cd} \psi \right) {- 5  \left( \bar{\psi} \Gamma_{cd} \psi X^c V^d + \bar{q} \Gamma_c \psi \bar{\psi} \Gamma^c \psi \right) }\right] = 0 \,,
\end{align}
due to the Fierz identities \eqref{usefulfierz11d} and \eqref{usefulfierz}. Observe that, in the case of the CE cohomology, the second cohomology class $\omega^{(6)}$ emerges once the 
first class $\omega^{(4)}_2$ is trivialized by adding the potential $A^{(3)}$, while here the second cohomology 
class $\omega^{(6)}$ emerges at the first stage of computation.\\
{Let us end this section by commenting on the absence of $2$-form cocycles: this is in agreement with the results in \cite{Figueroa-OFarrill:2015rfh,Figueroa-OFarrill:2015tan,Santi:2019kpx}. Nontrivial Spencer cohomologies, and therefore possible different supergravity backgrounds, could instead be obtained by considering different subalgebras in the definition of the filtered deformation.}

\section{Conclusions}\label{sec9}

In this paper we have shown that the Molien-Weyl integral formula, which allows us to derive invariant polynomials and, consequently, cohomologies, is a powerful mathematical tool to construct FDAs, simplifying the study of the underlying vacuum structure of supergravities.
After introducing the method, we have provided several applications, by considering diverse spacetime dimensions and amounts of supersymmetry, both in the case of flat superspaces and of curved supermanifolds. The results are displayed in Table \ref{tablesummary}. 

\begin{table}[h!]
\renewcommand{\arraystretch}{1.8}
\footnotesize
\centering
\begin{tabular}{ |p{3.5cm}|p{2.5cm}|p{3.5cm}|p{3.5cm}| }
 \hline
 \ \ \ \ \ \ \ \ \ \ Spacetime & \ Supersymmetry & \ \ \ \ \ Symmetry group & \ \ \ \ \ \ \ \ \  Cocycles \\ 
 \hline\hline
 \ \ \ \ \ \ \ \ \multirow{6}{*}{$D=4$ Flat}  &\ \ \ \ \ \ \  $N=1$ & \ \ \ \ \ \ \ \ \ \ \ \  $\mathfrak{so}(4)$ & $\omega^{(3)}$, $\omega^{(4)}_1$, $\omega^{(4)}_2$, $\ldots$  \\
 \cline{2-4}
   & \ \ \ \ \ \ \ \multirow{3.3}{*}{$N=2$} & \ \ \ \ \ \ \ \  $\mathfrak{so}(4)\oplus\mathfrak{u}(1)$   &$\omega^{(3)A}{}_B$, $\ldots$\\\cline{3-4}
  & & \ \ \ \ \ \ \,  $\mathfrak{so}(4)\oplus\mathfrak{su}(2)$ &  $\omega^{(2)}_1$, $\omega^{(2)}_2$, $\omega^{(3)}$\\\cline{3-4}
  &  &  \ \ \ \ \ \ \, $\mathfrak{so}(4)\oplus\mathfrak{u}(2)$& \\
  \cline{2-4}
  & \ \ \ \ \ \ \ \multirow{2.2}{*}{$N=4$}  & \ \ \ \ \ \ \ \ \ \ \ \ $\mathfrak{so}(4)$& $\omega^{(2)}_1$, $\omega^{(2)}_2$, $\ldots$\\\cline{3-4}
  & &  \ \ \ \ \ \,  $\mathfrak{so}(4)\oplus\mathfrak{su}(4)$& \\
  \hline
 \ \ \ \ \ \ \multirow{3.3}{*}{$D=4$ Curved}  & \ \ \ \ \ \ \ $N=1$ & \ \ \ \ \ \ \ \ \ \ \ \ $\mathfrak{so}(4)$ &$\omega^{(4)}$, $\omega^{(7)}$\\ 
 \cline{2-4}
  & \ \ \ \ \ \ \ $N=2$ &\ \ \ \ \ \,  $\mathfrak{so}(4)\oplus\mathfrak{so}(2)$ & $\omega^{(2)}$, $\omega^{(4)}$, $\omega^{(7)}$\\
 \cline{2-4}
  & \ \ \ \ \ \ \ $N=4$ & \ \ \ \ \ \, $\mathfrak{so}(4)\oplus\mathfrak{so}(4)$ &$\omega^{(4)}_1$, $\omega^{(4)}_2$, $\omega^{(4)}_3$, $\omega^{(7)}$\\
 \hline
 \ \ \ \ \ \ \ \ \multirow{3.3}{*}{$D=6$ Flat}  & \ \ \ \ \ $N=(4,0)$ &\ \ \ \ \ \,  $\mathfrak{so}(6)\oplus\mathfrak{usp}(4)$ & \\
 \cline{2-4}
 & \ \ \ \ \ $N=(2,2)$ &\ \ $\mathfrak{so}(6)\oplus\mathfrak{su}(2)\oplus\mathfrak{su}(2)$ &$\omega^{(3)}$\\
 \cline{2-4}
 & \ \ \ \ \ $N=(2,0)$ &\ \ \ \ \ \,  $\mathfrak{so}(6)\oplus\mathfrak{su}(2)$ &$\omega^{(3)}$\\
 \hline
 \ \ $D=10$ Flat. Type I & \ \ \ \ \ \ \ {$N=1$} & \ \ \ \ \ \ \ \ \ \ \ {$\mathfrak{so}(10)$} &{$\omega^{(3)}$, $\omega^{(7)}$, $\omega^{(9)}$, $\ldots$}\\
 \hline
 \,$D=10$ Flat. Type IIA & \ \ \ \ \ \ \ {$N=2$} &\ {$\mathfrak{so}(10)\oplus{u}(1)_L\oplus\mathfrak{u}(1)_R$} &{$\omega^{(2)}$, $\omega^{(3)}$, $\omega^{(4)}$, $\ldots$}\\
 \hline
 \,$D=10$ Flat. Type IIB & \ \ \ \ \ \ \ {$N=2$} & \ \ \ \ \ \ \ \ \ \ \ {$\mathfrak{so}(10)$} &{$\omega^{(3)|(AB)_0}$, $\omega^{(5)}$, $\omega^{(7)}$, $\ldots$} \\
 \hline
 \ \ \ \ \ \ \ \ $D=11$ Flat & \ \ \ \ \ \ \ $N=1$ & \ \ \ \ \ \ \ \ \ \ \ $\mathfrak{so}(11)$ &$\omega^{(4)}$, $\omega^{(7)}$\\
 \hline
 \ \ \ \ \ \ \ \ $D=12$ Flat & \ \ \ \ \ \ \ $N=1$ & \ \ \ \ \ \ \ \ \ \ \ $\mathfrak{so}(12)$ & $\omega^{(4)}_1$, $\omega^{(4)}_2$, $\ldots$\\
\hline
\end{tabular}  
\caption{\label{tablesummary} Cocycles in diverse dimensions and for different amounts of supersymmetry. Empty boxes indicate that the only nontrivial cohomologies are the constants.}
\end{table}
In some cases, the Hilbert-Poincaré series implies that the corresponding FDA must be infinite dimensional. We proposed some interpretations, but a complete comprehension is still missing and will deserve further research.
We also plan to extend our study to gauge hierarchies \cite{menoMario,CederPalm1,CederPalm2,Papadopoulos} in the context of deformations and to the framework of integral forms \cite{if1,if2,if3,if4,if5}.\\
We argue that the Molien-Weyl integral method could also be applied to write the invariant polynomials in the case in which the base space is enlarged by the presence of extra bosonic and fermionic fields. In particular, the latter may turn out to be crucial ingredients to trivialize FDAs, yielding the hidden superalgebra(s) underlying $D\geq 4$ supergravities in vacuum, in the spirit of \cite{DFd11, FDAnew1, FDAnew2,Ravera:2021sly}. Such study will be the subject of future endeavours. 

\section*{Acknowledgments}

{We are grateful to L. Andrianopoli, B. L. Cerchiai, R. D'Auria, and M. Trigiante for the illuminating discussions and suggestions.} \\
P. A. G. is partially supported by reseach funds of Univerisit\`a del Piemonte Orientale.
C. A. C. acknowledges GA\v CR grant EXPRO 19-28628X for financial support. R. N. acknowledges GA\v CR grant EXPRO 20-25775X for financial support. L. R. would like to thank the Department of Applied Science and Technology of the Polytechnic of Turin and the INFN for financial support.

\end{document}